\documentclass[a4paper,11pt]{article}
\usepackage{jheppub} 
\usepackage{lineno}
\usepackage{soul}
\usepackage[normalem]{ulem}
\usepackage{accents}
\usepackage{bbold}
\usepackage{xcolor}

\title{\boldmath  Quantum charged black holes}

   \author[a]{Yiji Feng,}
\author[b]{Hao Ma,}   
\author[c,d]{Robert B. Mann,}
  \author[a]{Yesheng Xue,}
    \author[a,c]{Ming Zhang  }
 \affiliation[a]{Department of Physics, Jiangxi Normal University,\\ Nanchang 330022, China}
 \affiliation[b]{Department of Physics, The Chinese University of Hong Kong, \\Hong Kong 999077, China}
 \affiliation[c]{Department of Physics and Astronomy, University of Waterloo,\\
Waterloo, Ontario N2L 3G1, Canada}
\affiliation[d]{Perimeter Institute for Theoretical Physics, \\ 31 Caroline St. N., Waterloo, Ontario N2L 2Y5, Canada}

\emailAdd{yijifeng@jxnu.edu.cn}
\emailAdd{haoma@link.cuhk.edu.hk}
\emailAdd{rbmann@uwaterloo.ca}
\emailAdd{yesheng-xue@jxnu.edu.cn}
\emailAdd{mingzhang@jxnu.edu.cn}

\abstract{Within the framework of braneworld holography, we construct a quantum charged black hole  localized on a three-dimensional anti-de Sitter (AdS) brane that intersects the asymptotic boundary of the four-dimensional AdS spacetime at the conformal defects and incorporates quantum backreaction effects from the conformal field theory (CFT) on the brane. This quantum charged black hole is an exact solution  of the semiclassical gravitational equation corresponding to a theory with higher curvature gravity and nonminimally coupled nonlinear gauge field. In the framework of double holography, we investigate the thermodynamics of the quantum charged black hole from three perspectives: a pure bulk perspective, in which four-dimensional classical  Einstein gravity couples to Maxwell electrodynamics and a 
codimension-one tensional brane; a brane perspective, where 
semiclassical higher curvature  gravity 
is subject to quantum backreaction from the holographic CFT on the brane, yielding a 
 quantum charged black hole; and a boundary perspective, where the defect CFT is coupled to a boundary CFT at the asymptotic boundary and  the degrees of freedom for defect quantum conformal matter is considered.  In so doing, we obtain doubly holographic formulations of both the first law of thermodynamics and the Smarr (energy) relations for the quantum charged black holes. }

\begin{document}
\maketitle
\flushbottom
\section{Introduction}

Without a theoretically self-consistent quantum theory of gravity, the classical Einstein equation can be extended to the semiclassical form  
\begin{equation}\label{dkljp983}
\mathrm{\mathbf{G}}\left( \mathrm{\mathbf{g}} \right)=\frac{8 \pi \mathrm{G}_d}{c^4}\left\langle \mathrm{\mathbf{T}}\left( \mathrm{\mathbf{g}} \right)\right\rangle\,,
\end{equation}
where $\mathrm{\textbf{G}}$ represents the Einsteinian curvature quantities associated with the bulk spacetime metric $\mathrm{\mathbf{g}}$, $G_d$ and $c$ are the $d$-dimensional Newton constant and the speed of light respectively, and $\left\langle \mathrm{\mathbf{T}}\left( \mathrm{\mathbf{g}} \right)\right\rangle$ denotes the expectation value of the renormalized stress-energy tensor of quantum fields. This equation encodes the backreaction or corrections of quantum matter on the classical geometry $\mathrm{\mathbf{g}}$. However, the equation is challenging to solve non-perturbatively. As pointed out in \cite{Emparan:1999wa,Emparan:1999fd,Emparan:2002px,Emparan:2020znc}, exact calculations of $\left\langle \mathrm{\mathbf{T}}\left( \mathrm{\mathbf{g}} \right)\right\rangle$ for the massless conformally coupled scalar field and its backreaction can only be attained in the three-dimensional cases \cite{Steif:1993zv,Lifschytz:1993eb,Shiraishi:1993qnr}, specifically for the Bañados-Teitelboim-Zanelli (BTZ) spacetime \cite{Banados:1992wn,Banados:1992gq} and also a two-dimensional model \cite{Callan:1992rs,Strominger:1994tn}, particularly the one in the well-known Jackiw-Teitelboim (JT) gravity \cite{Jackiw:1984je,Teitelboim:1983ux,Almheiri:2014cka,Maldacena:2016upp,Engelsoy:2016xyb}. In other cases where $d\geq 4$, $\left\langle \mathrm{\mathbf{T}}\left( \mathrm{\mathbf{g}} \right)\right\rangle$ can only be obtained perturbatively, making it difficult to analyze its backreaction effects.

In \cite{Randall:1999vf}, a Randall-Sundrum (RS) geometry was constructed within a non-compact five-dimensional anti-de Sitter (AdS) bulk spacetime. This construction involves embedding a four-dimensional brane (or a three-brane) with a finely tuned positive tension. As a result,   four-dimensional Newtonian gravitational effects, as well as   low-energy and long-distance effects of Einstein gravity, can be mimicked through the normalizable zero mode and the Kaluza-Klein mode.  Following this, the Karch-Randall (KR) brane theory was tested   \cite{Karch:2000ct} for the localization of  AdS gravity on the brane, even in the presence of a divergent zero mode wave function. Recently JT gravity was shown to be derivable from the KR braneworld by considering small fluctuations of the brane \cite{Geng:2022slq,Geng:2022tfc}. RS branes are codimension-one Minkowski or de Sitter (dS)  branes embedded in the ambient $(d+1)$-dimensional AdS${}_{d+1}$ spacetime, whereas KR branes are codimension-one AdS${}_{d}$ branes embedded in the ambient AdS${}_{d+1}$ space \cite{Geng:2022slq,Randall:1999ee,Randall:1999vf}. These brane theories offer intriguing prospects for braneworld scenarios \cite{Garriga:1999yh,Geng:2023qwm,Geng:2023iqd,Geng:2024xpj} and have significant implications for the realization of the holographic principle \cite{tHooft:1993dmi,Susskind:1994vu,Maldacena:1997re,Gubser:1998bc,Witten:1998qj,deHaro:2000wj}. In the AdS${}_{d+1}$ holography framework, asymptotically AdS${}_{d+1}$ classical gravity couples to a codimension-one AdS${}_{d}$ brane. This brane  intersects the asymptotic boundary of the AdS${}_{d+1}$ spacetime at conformal defects where the DCFT${}_{d-1}$ (DCFT${}_{d-1}$) \cite{Yamaguchi:2002pa} theory resides.  On the asymptotic boundary, there is a  boundary CFT${}_{d}$ (BCFT${}_{d}$) \cite{Cardy:2004hm,McAvity:1995zd,Takayanagi:2011zk,Fujita:2011fp,Geng:2022dua}  that is coupled with the DCFT${}_{d-1}$ \cite{Jensen:2013lxa} on the intersection points of the brane and the asymptotic boundary. Furthermore, in the braneworld holography framework \cite{Randall:1999ee,Randall:1999vf,Karch:2000ct,Karch:2001cw}, there is an asymptotically AdS${}_{d}$ semiclassical gravity on the brane that  is coupled with a holographic CFT${}_{d}$ that  communicates with the BCFT${}_{d}$ at the half-space of the asymptotic boundary of the AdS${}_{d+1}$ spacetime. This scenario is known as double holography \cite{Karch:2000gx,Almheiri:2019hni,Chen:2020uac,Karch:2022rvr}: classical gravity in AdS${}_{d+1}$ corresponds to localized quantum gravity on the AdS${}_{d}$ KR brane, and further this (semiclassical) quantum gravity on the brane is holographically dual to the DCFT${}_{d-1}$ at the defect via AdS holography \cite{Neuenfeld:2021wbl}. This formulation has been used to study quantum extremal surfaces \cite{Hubeny:2007xt,Faulkner:2013ana,Lewkowycz:2013nqa,Engelhardt:2014gca,Penington:2019npb,Almheiri:2019psf,Geng:2020qvw,Almheiri:2020cfm,Chen:2020hmv,Geng:2020fxl,Chen:2020jvn,Geng:2021mic,Ling:2020laa,Grimaldi:2022suv,Chang:2023gkt,Myers:2024zhb} beyond holographic entanglement entropy \cite{Ryu:2006bv,Ryu:2006ef} for the information paradox \cite{Hawking:1975vcx,Hawking:1976ra,Page:1993wv,Page:2013dx}.

Some time ago, a quantum BTZ (quBTZ) black hole was constructed \cite{Emparan:1999wa,Emparan:1999fd}. This was based on a two-brane scenario contained within a four-dimensional bulk, diverging from the original RS scenario. The introduced brane deformed the semiclassical equation \eqref{dkljp983} by incorporating additional higher curvature corrections to Einstein gravity. These corrections originate from a spatial cutoff of the brane, leaving the original complete three-dimensional CFT (CFT${}_3$) at the asymptotic boundary dual to the AdS${}_4$ bulk to now consist of a codimension-one brane and an $S^2 / \mathbb{Z}_2$ half CFT${}_3$ boundary \cite{Karch:2000ct}. Recently in \cite{Emparan:2020znc} the same static quBTZ black hole was further explored by considering the mechanisms of backreaction and higher curvature corrections. This successful construction of the three-dimensional quBTZ black hole on the brane is a realization of the  braneworld holography principle: gravity can emerge from the brane. The reason we refer to the black hole on the brane as a quantum black hole is that it is a solution of the deformed semiclassical gravitational field equation \cite{Emparan:2002px}.

There are many intriguing aspects of the quBTZ black hole on the brane \cite{Emparan:1999wa,Emparan:1999fd,Emparan:2020znc,Frassino:2022zaz}. First and foremost, the quBTZ black hole is derived from the four-dimensional C-metric in an AdS${}_4$ background, a metric that describes a uniformly accelerating black hole \cite{Plebanski:1976gy}. The reason for using the C-metric is that a black hole on a brane in AdS should be accelerating; this can be realized via the conical singularity of the  four-dimensional accelerating black hole \cite{Emparan:1999wa,Podolsky:2000pp,Dias:2002mi}. Utilizing the braneworld construction, the relationship between the brane tension and its position is constrained by the Israel junction condition \cite{Israel:1966rt}. This means the junction condition can be satisfied at a specific position \cite{Kudoh:2004ub}. The quBTZ black hole exhibits BTZ-like characteristics, and its mass -- determined by the asymptotic deficit angle on the brane -- matches the effective mass of the four-dimensional bulk, as derived from the thermodynamic first law relation \cite{Emparan:1999fd}. This model accurately incorporates the backreaction of the cutoff holographic CFT and higher curvature corrections to the localized gravity on the brane, both of which originate from integrating out the ultraviolet (UV) degrees of freedom of the CFT at the asymptotic boundary. The introduction of the brane into the bulk fundamentally alters the thermodynamics of the accelerating black hole, resulting in a thermodynamic first law that differs from those found in \cite{Appels:2016uha,Abbasvandi:2018vsh,Gregory:2019dtq,Appels:2017xoe,Anabalon:2018ydc}.

Recently, significant progress has been made in the study of holographic quantum black holes on the KR brane. The static quBTZ black hole was extended to include rotation \cite{Emparan:2020znc}, thus becoming stationary. The renormalized CFT stress-energy tensor was obtained and the holographic quantum entropy of this black hole were shown to satisfy the thermodynamic first law. In the limit of vanishing backreaction, the solution can be reduced to either the rotating BTZ black hole or a rotating conical defect. Subsequently, by starting from the AdS${}_4$ C-metric and setting the AdS${}_3$ radius $\ell_3$ to $i R_3$ (with $R_3$ defined as the radius of the dS brane with a positive cosmological constant; the corresponding three-dimensional effective cosmological constant can be attained by considering a brane with large enough tension \cite{Karch:2000ct}), it was shown that one can derive the quantum dS black hole, with or without rotation \cite{Emparan:2022ijy,Panella:2023lsi}. Remarkably, while no classical black hole exists in a three-dimensional spacetime, within the braneworld holography framework, a horizon emerges due to the backreaction of the quantum conformal fields 
(the CFT degrees of freedom) 
residing on the brane. Beyond the many potential research topics highlighted in \cite{Emparan:2020znc,Emparan:2022ijy,Panella:2023lsi}, there have been developments in the field of quantum black holes, particularly for the quBTZ black hole. These developments partially encompass holographic complexity \cite{Emparan:2021hyr,Chen:2023tpi}, black hole chemistry \cite{Frassino:2022zaz,Johnson:2023dtf,Frassino:2023wpc,HosseiniMansoori:2024bfi,Wu:2024txe}, inner structure \cite{Kolanowski:2023hvh}, and quantum inequalities \cite{Frassino:2024bjg}. These investigations enrich our understanding of the holographic and thermodynamic properties of quantum black holes on the AdS${}_3$ brane.

Electromagnetic fields play a prominent role in spacetime structure \cite{Emparan:2020znc,Panella:2023lsi}, influencing properties such as singularities  \cite{Ayon-Beato:1998hmi,Cardoso:2017soq,Kolanowski:2023hvh}, thermodynamic characteristics \cite{Kastor:2009wy,Cvetic:2010jb,Dolan:2011xt,Kubiznak:2012wp,Wei:2019uqg,Kubiznak:2016qmn,Xiao:2023lap,Wei:2023mxw}, and many other aspects \cite{Lanir:2018vgb,Hollands:2019whz,McMaken:2023tft,Cannizzaro:2024yee}. It is therefore both natural and necessary to explore a quantum charged black hole in the KR braneworld context. The feasibility of this exploration is enhanced by the availability of the AdS charged  C-metric \cite{Kinnersley:1970zw}, which is in an  appropriate form to serve as a starting point for studying a three-dimensional quantum charged black hole. 

In this paper, we will demonstrate  that a quantum charged black hole can be obtained within the framework of braneworld holography. 
 It is shown the quantum charged black hole to be  quite different from the charged BTZ black hole  \cite{Martinez:1999qi,Chan:1994qa}, resembling more closely to the Reissner-Nordström (RN) AdS black hole  in
 terms of the form of the metric function and the associated gauge field. 
In the next section, we will give a brief review of the charged AdS C-metric. In section \ref{wjopi2}, we will present the explicit form of the three-dimensional quantum charged black holes on the KR brane. The holographic stress-energy tensor encoding the backreaction of the quantum CFT${}_3$ on the brane will be studied. We will also calculate thermodynamic quantities related with the quantum charged black holes. In section \ref{dfjj39882}, we will study the thermodynamics of the quantum charged black holes within the double holography framework.  We obtain the  doubly holographic formulations of both 
the first law of thermodynamics 
and the Smarr (energy) relations, generalizing previous results for
holographic black hole chemistry
\cite{Cong:2021fnf,Frassino:2022zaz,Ahmed:2023snm}. 
The final section will be devoted to closing remarks. Throughout the paper, we will set $c=1$ for convenience. Additionally, the symbols used will be consistent with those in \cite{Emparan:2020znc}; please refer to the symbol glossary in Appendix A of \cite{Emparan:2020znc} for clarification.

\section{A brief review of the charged AdS C-metric}

We will first derive an asymptotically AdS charged C-metric solution in a specific form through transformations of coordinates and rescalings of parameters. We will then analyze the ranges of the parameters for the charged AdS C-metric solution.

\subsection{Charged C-metric solutions}

As a member of the Plebański-Demiański family of type-D metrics \cite{Plebanski:1976gy}, the study of the C-metric has a long history \cite{Weyl:1917gp,levi1918t,newman1961new,robinson1962robinson}. In 1970, the AdS  charged C-metric solution was obtained in the form \cite{Kinnersley:1970zw}
\begin{equation}\label{oij38934}
\mathrm{d} s^2=\frac{1}{A^{2}(x-y)^{2}}\left(-\mathfrak{F} \mathrm{d}\tilde{t}^2+\mathfrak{F}^{-1} \mathrm{d} y^2+G^{-1} \mathrm{d} x^2+G \mathrm{d} \phi^2\right)\,,
\end{equation}
where
\begin{align}
    \mathfrak{F}(y)&=A^2 e^2 y^4+2 A m y^3-k y^2+\lambda\,,\\
     G(x)&=-A^2 e^2 x^4-2 A m x^3+k x^2+1
\end{align}
with $A\geq 0, m\geq 0, e\geq 0$ being the acceleration, mass, and electric charge parameters, respectively. The discrete values of $k$ are $0,\,\pm 1$. $\lambda\geq -1$ is related to the cosmological constant. The C-metric can be recovered with $k=-1,\, \lambda\to -1$ and for $\lambda>-1$ we have the AdS C-metric. In the limit of the static black hole with $A\to 0$,\, $k=-1$ refers to a spherical horizon, while $k=0, 1$ refer to $\mathbb{R}^2$ flat and hyperbolic horizons, respectively \cite{Mann:1996gj}.    It is straightforward to verify that the solution \eqref{oij38934} satisfies the classical gravitational field equation 
\begin{equation}\label{koiejpeo4}
R_{ab}+\frac{1}{2} \underaccent{\bar}{F}_{c d}  \underaccent{\bar}{F}^{c d}  g_{a b}+2 \underaccent{\bar}{F}_a{}^{c} \underaccent{\bar}{F}_{c b}+\frac{3}{\ell_4^2} g_{ab}=0\,, \quad \ell_4=\frac{1}{A \sqrt{\lambda+1}}\,,
\end{equation}
where $\ell_4$ is the AdS${}_4$ radius of the spacetime, and $\mathbf{\underaccent{\bar}{F}}$  is the electromagnetic field tensor satisfying $\nabla\cdot\mathbf{\underaccent{\bar}{F}}=0$, where
\begin{equation}\label{jireoj39}
\mathbf{\underaccent{\bar}{F}}=\mathrm{d}\underaccent{\bar}{\mathcal{A}}\,,\quad\underaccent{\bar}{\mathcal{A}}_a=e y (\mathrm{d} \tilde{t})_a
\end{equation}
with $\underaccent{\bar}{\mathcal{A}}$ the associated gauge potential \cite{Hong:2003gx}.

We now carry out coordinate transformations and parameter rescalings on the C-metric \eqref{oij38934} by \cite{Emparan:2020znc}
\begin{equation}\label{dkjaoe834}
\begin{aligned}
\lambda&=\left(\ell / \ell_3\right)^2\,, \quad A=1 / \ell\,, \quad k=-\kappa\,, \quad 2 m A=\mu\,, \\ \quad y&=-\ell / r\,,\quad \tilde{t} = t / \ell \, ,
\end{aligned}
\end{equation}
where $\mu$ encodes the mass of the black hole, $\ell$ is the inverse of the black hole acceleration, $\ell/\ell_3$ replaces the cosmological constant parameter $\lambda$,  $\kappa$ (like $k$) also parameterizes the horizon topology of the black hole, the coordinate $r$ is used to represent the coordinate $y$, and the coordinate $\tilde{t}$ is rescaled by the inverse acceleration parameter $\ell$. 
We can also transform the electric charge parameter $e$ to a rescaled electric charge parameter $q$ by   $ q=eA$. Since $A\geq 0,\, \ell \geq 0$,
  setting  $\ell$ to zero  corresponds to $A\to\infty$. As a result, we obtain a new form for the metric \eqref{oij38934} and the electromagnetic potential \eqref{jireoj39}, given by 
\begin{equation}\label{j9qj398}
\mathrm{d} s^2=\frac{1}{\Omega^2}\left[-H(r) \mathrm{d} t^2+\frac{ \mathrm{d} r^2}{H(r)}+r^2\left(\frac{ \mathrm{d} x^2}{G(x)}+G(x) \mathrm{d}\phi^2\right)\right]\,,
\end{equation}
\begin{equation}\label{jsp39}
\mathbf{\underaccent{\bar}{F}}=\mathrm{d}\underaccent{\bar}{\mathcal{A}}\,,\,\quad \underaccent{\bar}{\mathcal{A}}_a=-\frac{q\ell}{r}\left(\mathrm{d} t\right)_a\,,
\end{equation}
where
\begin{eqnarray}
    \Omega &=&1+\frac{xr}{\ell}\,,\\
    H(r)&=&\frac{r^2}{\ell_3^2}+\kappa-\frac{\mu \ell}{r}+\frac{q^2 \ell^2}{r^2}\,,\label{jioperjf83}\\
    G(x)&=&1-\kappa x^2-\mu x^3 -q^2 x^4\,.
\end{eqnarray}
This metric satisfies the Einstein equation
\begin{equation}\label{rejop3988}
R_{a b}+2 \underaccent{\bar}{F}_a{}^{c} \underaccent{\bar}{F}_{c b}+\frac{1}{2} \underaccent{\bar}{F}_{c d}  \underaccent{\bar}{F}^{c d}  g_{a b}=-3\left(\frac{1}{\ell^2}+\frac{1}{\ell_3^2}\right) g_{a b}\equiv -\frac{3}{\ell_4^2}g_{ab}\,,
\end{equation}
where 
\begin{equation}\label{dklaje292}
\ell_4=\left(\frac{1}{\ell^2}+\frac{1}{\ell_3^2}\right)^{-1 / 2} 
\end{equation}
is the $\mathrm{AdS}_4$ length scale of the bulk black hole spacetime. Note that in the limit $\ell\to\infty$, i.e., $A\to 0$, we have $\ell_4=\ell_3$. The rescaling of the electric charge parameter $e$ to $q$ makes $qx$ dimensionless and  will facilitate our calculations in what follows.

\subsection{Parameter Ranges }

A basic requirement for the ranges of the parameters of the black hole \eqref{j9qj398} is that the signature remains invariant in the domain of outer communication. The inner and outer horizons $r_{\mp}$ of the black hole are determined by $H(r_\mp)=0$. The conformal boundary $r_c$ is determined by $\Omega=0$. $H(r)$ and $G(x)$ must not change sign between $r_+<r<r_c$.

For the angular directions, the criteria for the range of $x$ are governed by the regularity at symmetry axes of $\partial_\phi$ \cite{Emparan:1999fd} and the requirement of $G(x)\geq 0$. For $x\to 0$ these criteria can be satisfied directly; for $\kappa=1$ we have  $x\leq 1$ in case of   $\mu=0, \,q=0$. In the general case of $\mu\neq 0, q\neq 0$, by continuity we have
\begin{equation}\label{kldfa489}
0 \leq x \leq x_1\,,
\end{equation}
where $x_1$ is the minimal positive solution of the equation 
\begin{equation}\label{io3qjrf843}
G(x)=0\,.
\end{equation}
Since \eqref{dkjaoe834} implies $\mu\geq 0$, we obtain
\begin{equation}
\mu=\frac{-q^2 x_1^4-\kappa  x_1^2+1}{x_1^3}\geq 0
\end{equation}
from \eqref{io3qjrf843}. We can further derive that
\begin{equation}\label{jeio4829}
x_1 \in(0,\,1)\,,\quad 0\leq q\leq \sqrt{\frac{1-\kappa x_1^2}{x_1^4}}\,,\quad  (\text{or}\;  x_1=1\,,\, q=0)\,,\quad \text { for } \;\kappa=+1 \, ,
\end{equation}
which indicates that the range of $x_1$ is finite for the spherical horizons. In contrast, we have $x_1 \in(0,\,\infty)$ for $\kappa=0,\,-1$.   Note that for all cases $x_1$ cannot be zero. Just as in the uncharged case \cite{Emparan:2020znc}, $\mu$ is a monotonically decreasing function of $x_1$; specifically, we have
\begin{equation}
\mu\to\infty\,, \quad\mathrm{for}\quad x_1\to 0\,.
\end{equation}
If $x_1$ and $q$ approach their maximal values in \eqref{jeio4829} then $\mu\to 0$.

\begin{figure}
    \centering
    \includegraphics[width=0.5\linewidth]{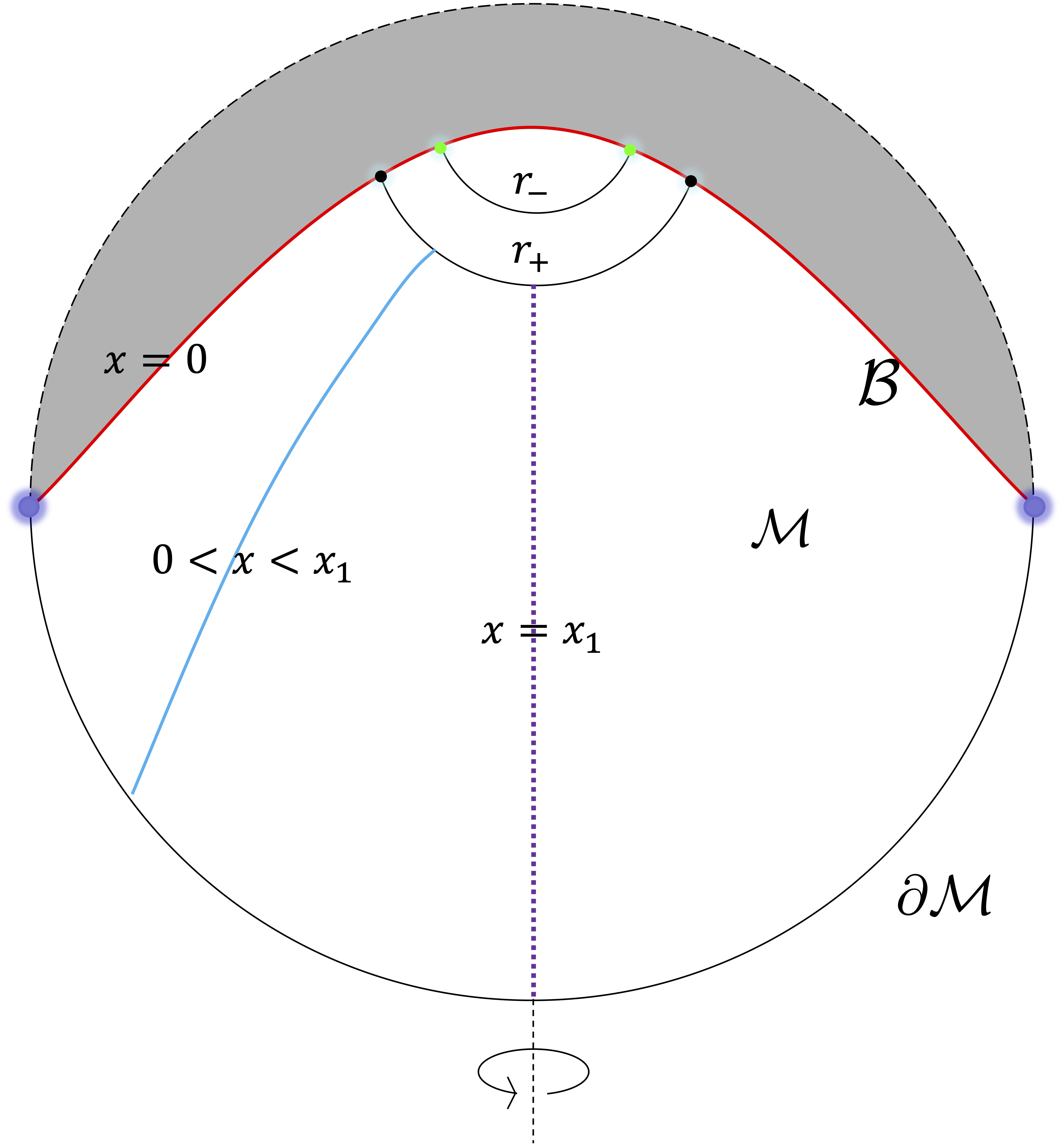}
    \caption{  The braneworld construction for the C-metric spacetime \eqref{j9qj398} represented by $\mathcal{M}$, shown
    at a slice of constant  $t$ and $\phi$. $\partial \mathcal{M}$ stands for the asymptotic boundary where the BCFT${}_3$ lives. A KR brane (red line) cutting off the bulk near the asymptotic boundary is placed at $x=0$; this is  tantamount to a UV cutoff of the asymptotic CFT. Localized three-dimensional gravity arises on the codimension-one brane $\mathcal{B}$. A bulk charged black hole  is also housed on the brane, with its corresponding inner and outer horizons $r_{\mp}$ represented by the green and black dots.  The dotted purple line denotes the $\phi$ axis $x=x_1$ and  a blue line traversing the bulk region for some $0<x<x_1$ is also shown.   We do not display the second patch here that is glued to the current one at the brane via the Israel junction conditions.   }
    \label{ji3o82ujhfn3pqo}
\end{figure}

\section{Quantum charged black holes}\label{wjopi2}

\subsection{Quantum charged black hole solutions}

With the four-dimensional AdS charged C-metric spacetime \eqref{j9qj398} at hand, we can obtain a three-dimensional quantum charged black hole by using the KR braneworld formulation by considering a configuration where an AdS${}_3$ brane with tension $\tau$ is placed in the bulk spacetime (see Fig. \ref{ji3o82ujhfn3pqo}). In the notations of \cite{Frassino:2022zaz,Climent:2024nuj}, the total action of the system can be written as
\begin{equation}\label{iurj389}
I=I_{\text {bulk }}[\mathcal{M}]+I_{\mathrm{GHY}}[\partial \mathcal{M}]+I_{\text {brane }}[\mathcal{B}]\,,
\end{equation}
where the bulk action, Gibbons-Hawking-York (GHY) action, and the action of the brane are respectively given by
\begin{eqnarray}
  I_{\text {bulk }}&=&\frac{1}{16 \pi \mathrm{G}_{4}} \int_{\mathcal{M}} \mathrm{d}^{4} x \sqrt{-g}\left(\mathcal{R}+\frac{6}{\ell_4^2}\right)- \frac{1}{4g_4^2}\int_{\mathcal{M}} \mathrm{d}^{4} x \sqrt{-g}F_{ab}F^{ab}\,,\label{j201dm}\\ 
  I_{\mathrm{GHY}}&=&\frac{1}{8 \pi \mathrm{G}_{4}} \int_{\partial \mathcal{M}} \mathrm{d}^3 x \sqrt{-h} K\,,\\
  I_{\text {brane }}&=&-\tau \int_{\mathcal{B}} \mathrm{d}^3 x \sqrt{-h}\,,
\end{eqnarray}
where $\mathcal{R}$ is the Ricci scalar of the bulk $\mathcal{M}$ with a metric $g_{ab}$ whose determinant is $g$, $K$ is the GHY extrinsic curvature scalar on the asymptotic boundary $\partial \mathcal{M}$, and $h$ is the determinant of the induced metric on the brane $\mathcal{B}$, and $\mathrm{G}_4$ is the four-dimensional Newton constant. The electromagnetic field tensor $\mathbf{F}$ is minimally coupled with the spacetime curvature through a dimensionless gauge coupling constant $g_4$ (it is denoted as $g_{\star}$ in \cite{Climent:2024nuj}), which determines the normalization of the gauge field and  is related with $\mathbf{\underaccent{\bar}{F}}$ in \eqref{jireoj39} or \eqref{jsp39} by
\begin{equation}\label{j2091}
 4\mathbf{\underaccent{\bar}{F}^2}=\ell_{\star}^2 \mathbf{F^2}   
\end{equation}
with $\ell_{\star}^2\equiv 16\pi \mathrm{G}_4/g_4^2$ \cite{Climent:2024nuj}.

In the KR braneworld model, two patches of the AdS${}_4$ geometries are glued together along the brane obeying the Israel junction condition \cite{Israel:1966rt}. The tension of the brane is \cite{Kudoh:2004ub,Emparan:2020znc}
\begin{equation}\label{fjioe3894}
\tau=\frac{1}{2 \pi \mathrm{G}_4 \ell}
\end{equation}
if we place the brane at $x=0$ in the bulk spacetime \eqref{j9qj398}. On the brane, we have a three-dimensional induced effective theory of gravity coupled to the gauge field, just as in the bulk.    The quantum charged black hole on the brane mapped from the classical C-metric \eqref{j9qj398} in the bulk is
\begin{equation}\label{djioee8349}
\mathrm{d} s^2=-H(r) \mathrm{d} t^2+\frac{\mathrm{d} r^2}{H(r)}+r^2 \mathrm{d} \phi^2\,,
\end{equation}
where $H(r)$ is given by \eqref{jioperjf83}. 

Due to the conical deficit of the metric \eqref{djioee8349}, the range of the coordinate $\phi$ is $[0, 2\pi \Delta]$ with\footnote{We can expand the metric \eqref{jioperjf83} at $x=x_1+\delta$ with $\delta\ll 1$ to see this \cite{Panella:2023lsi}.}
\begin{equation}
\Delta=\frac{2}{\left|G^{\prime}\left(x_1\right)\right|}=\frac{2 x_1}{3-\kappa x_1^2+q^2 x_1^4}\,.
\end{equation}
However, we can set the azimuthal coordinate $\phi$ to   have the canonical range $[0, 2\pi ]$ via the coordinate transformations
\begin{equation}\label{jeioe3}
t=\Delta \bar{t}\,,  \quad r=\frac{\bar{r}}{\Delta}\,,\quad \phi=\Delta \bar{\phi}\,,
\end{equation}
which result in a canonical form of the metric \eqref{djioee8349} for the quantum charged black hole as
\begin{equation}\label{iroejreop1093}
\mathrm{d} s^2=-\bar{H}(\bar{r}) \mathrm{d} \bar{t}^2+\frac{\mathrm{d} \bar{r}^2}{\bar{H}(\bar{r})}+\bar{r}^2 \mathrm{d} \bar{\phi}^2\,,
\end{equation}
where  we can now identify $\bar{\phi}$ with $\bar{\phi}+2\pi$. Furthermore, the metric function $\bar{H}(\bar{r})$ is
\begin{equation}\label{j923j9}
\bar{H}(\bar{r})=\frac{\bar{r}^2}{\ell_3^2}-8 \mathcal{G}_3 M-\frac{\ell F(M)}{\bar{r}}+\frac{\Delta ^4 q^2 \ell^2}{\bar{r}^2}\,,
\end{equation}
where $M$ is the mass of the black hole, $\ell_3$ (first introduced in \eqref{dkjaoe834}) is the $\mathrm{AdS}_3$ radius on the brane, and
\begin{equation}\label{eajj38}
\mathcal{G}_3=\frac{1}{2 \ell} \mathrm{G}_4 
\end{equation}
is a three-dimensional `renormalized' Newton constant. 
The function $F(M)$ reads
\begin{equation}\label{398ur82jemd}
F(M)=\mu \Delta^3 \geq 0\,,
\end{equation}
where 
\begin{equation}\label{joepj398}
M=-\frac{\kappa}{8 \mathrm{G}_3} \frac{\ell}{\ell_4} \Delta^2 
\end{equation}
expresses the    mass   in terms of the conical deficit \cite{Emparan:1999fd,Emparan:2020znc,Kudoh:2004ub}. The gauge potential \eqref{jsp39} becomes
\begin{equation}\label{djkoeje43}
   \bar{\mathcal{A}}_\mu \mathrm{d}\bar{x}^\mu=-\frac{2q\ell \Delta^2}{\ell_{\star}\bar{r}}\mathrm{d}\bar{t}\,.
\end{equation}
 Solving  $\bar{H}(\bar{r})=0$ yields the outer event horizon $\bar{r}_+$ and inner Cauchy horizon $\bar{r}_-$  of the quantum charged black  hole on the brane with $\bar{r}_{\pm}=\Delta r_{\pm}$. 

Comparing the induced metric function $\bar{H}(\bar{r})$ given by \eqref{j923j9} and the gauge potential $\bar{\mathbf{\mathcal{A}}}$ from  \eqref{djkoeje43} with those of the  RN-AdS black hole and the charged BTZ black hole \cite{Martinez:1999qi,Chan:1994qa}, we see that the quantum charged black hole is quite similar to the RN-AdS black hole in form. Unlike the BTZ black hole,  terms logarithmic in the radial variable 
do not appear.   In the following subsection, we will explain this by showing that a nonlinear electromagnetic field arises on the brane.

\subsection{Quantum backreactions}

The metric \eqref{iroejreop1093} and the gauge potential  \eqref{djkoeje43} for the quantum charged black hole  solves a deformed semiclassical gravitational field equation
\begin{equation}\label{dkljp9jidk83}
\bar{\mathbf{G}}\left( \mathrm{\mathbf{g}} \right)=8 \pi \mathrm{G}_4\left\langle \bar{\mathbf{T}}\left( \mathrm{\mathbf{g}} \right)\right\rangle,
\end{equation}
where $\bar{\mathbf{G}}$ equals the Einstein term $\mathrm{\mathbf{G}}\left( \mathrm{\mathbf{g}} \right)$ in \eqref{dkljp983} together with higher curvature correction terms on the brane, and $\left\langle \bar{\mathbf{T}}\left( \mathrm{\mathbf{g}} \right)\right\rangle$ encodes the extra leading order contributions from the holographic CFT${}_3$ on the brane. Using the notation of \cite{Frassino:2022zaz}, the field equation \eqref{dkljp9jidk83} can be viewed as being derived from the low-energy effective action $I$ as
\begin{equation}\label{idj389}
I=I_{\text {Bgrav }}[\mathcal{B}]+I_{\mathrm{CFT}}[\mathcal{B}]\,,
\end{equation}
where the first term on the right side of \eqref{idj389} 
\begin{equation}\label{jeiuj3389}
\begin{aligned}
I_{\text {Bgrav }}=& \frac{\ell_4}{8 \pi \mathrm{G}_4} \int \mathrm{d}^3 x \sqrt{-h}\left[\frac{4}{\ell_4^2}\left(1-\frac{\ell_4}{\ell}\right)+R+\ell_4^2\left(\frac{3}{8} R^2-R_{a b} R^{a b}\right)+\cdots\right]\\&+\int \mathrm{d}^3 x \sqrt{-h} f\left(\widetilde{F}_{ab},\, \nabla_a,\,R_{abcd} \right)
\end{aligned}
\end{equation}
is the gravitational contribution on the two-brane. Here  $R$ is the Ricci scalar on the brane, the covariant derivative $\nabla_a$ is compatible with the metric $h_{ab}$ on the brane,  $\widetilde{\mathbf{F}}=\mathrm{d}\bar{\mathcal{A}}$ is the electromagnetic tensor on the brane related to the gauge potential \eqref{djkoeje43} and $f(\widetilde{F}_{ab},\, \nabla_a,\,R_{abcd} )$ is a function of the electromagnetic field tensor, the spacetime curvature, and their derivatives. \footnote{
Shortly after our paper appeared, the necessity of the inclusion of electromagnetic-curvature coupling terms such as $R \mathbf{\widetilde{F}}^2$ was pointed out in \cite{Climent:2024nuj} based on the result in \cite{Taylor:2000xw}, which implies there will be current on the brane.} The second term on the right side of \eqref{idj389} is the action of the holographic CFT${}_3$ on the brane whose explicit expression cannot be given in   closed form but can be holographically defined by the bulk  \cite{Emparan:2020znc}.

We note that the Lagrangian of the electromagnetic field on the brane is no longer simply $\widetilde{\mathbf{F}}^2$, and a  non-linear electromagnetic field nonminimally coupled with three-dimensional induced effective gravity arises \cite{Climent:2024nuj}. It can be verified that if we set $f=\widetilde{\mathbf{F}}^2$, the equation of motion for the electromagnetic field cannot be satisfied by the gauge potential \eqref{djkoeje43}. The non-linearity emerges because the gravitational theory on the brane is modified by higher curvature terms; in other words, the metric \eqref{iroejreop1093} and gauge potential \eqref{djkoeje43} on the brane are induced from \eqref{j9qj398} and \eqref{jsp39} in the bulk, respectively, and the former equations of motion for the fields in the bulk are not obeyed anymore. As pointed out in \cite{Climent:2024nuj}, we explicitly have the scalar functional for the nonlinear electromagnetic field on the brane as \cite{Taylor:2000xw}
\begin{equation}\label{jfow329}
f\left(\widetilde{F}_{ab},\, \nabla_a,\,R_{abcd} \right)=-\frac{5\ell_4}{8g_4^2}\widetilde{\mathbf{F}}^2+\frac{2 \ell_4^3}{g_4^2}\bar{f}\,,
\end{equation}
where
\begin{equation}
\bar{f}=\frac{1}{288} R \mathbf{\widetilde{F}}^2-\frac{5}{8} R^a_b \widetilde{F}_{a c} \widetilde{F}^{b c}+\frac{5}{24}\left(\nabla\cdot \widetilde{\mathbf{F}}\right)^2+\frac{3}{98} \widetilde{F}^{a b}\left(\nabla_b \nabla^c \widetilde{F}_{c a}-\nabla_a \nabla^c \widetilde{F}_{ cb}\right).
\end{equation}

According to the low-energy effective action \eqref{jeiuj3389} for the brane, we can define some three-dimensional quantities. In \eqref{jeiuj3389}, an effective three-dimensional Newton constant $\mathrm{G}_3$ can be identified as
\begin{equation}\label{dejoiefjre}
\mathrm{G}_3=\frac{1}{2 \ell_4} \mathrm{G}_4=\frac{\ell}{\ell_4}\mathcal{G}_3\,,
\end{equation}
where in the last step we have used \eqref{eajj38}. Following \cite{Emparan:2020znc}, an AdS${}_3$ length scale $L_3$ on the brane can be extracted as
\begin{equation}\label{jeo3i28}
\frac{1}{L_3^2}=\frac{2}{\ell_4^2}\left(1-\frac{\ell_4}{\ell}\right)=\frac{1}{\ell_3^2}\left[1+\frac{\ell^2}{4 \ell_3^2}+\mathcal{O}\left(\frac{\ell^4}{\ell_3^4}\right)\right]
\end{equation}
upon using \eqref{dklaje292} and neglecting terms quadratic in  $\ell^2$.   It is evident that $L_3=\ell_3$ when $\ell\to 0$, which means that the backreaction from the CFT${}_3$ on the brane vanishes,  or the brane approaches the $\mathrm{AdS}_4$ boundary.  In contrast, for $\ell\to\infty$, which  by \eqref{fjioe3894} implies that the tension of the brane vanishes, we have $\ell_3\to\ell_4$ and $\ell_4\to \sqrt{2}L_3$. Besides, we can define a three-dimensional gauge coupling constant $g_3$ as
\begin{equation}
g_3^2=\frac{2 g_4^2}{5 \ell_4}\,,
\end{equation}
such that \eqref{jfow329} can be transformed as 
\begin{equation}\label{jf291z}
f\left(\widetilde{F}_{ab},\, \nabla_a,\,R_{abcd} \right)=-\frac{1}{4g_3^2}\widetilde{\mathbf{F}}^2+\frac{4 \ell_4^2}{5g_3^2}\bar{f}\,.
\end{equation}
With Eqs. \eqref{dejoiefjre}-\eqref{jf291z}, the brane action \eqref{jeiuj3389} can be recast as
\begin{equation}\label{jeiuj3e389}
\begin{aligned}
I_{\text {Bgrav }}=&\frac{1}{16 \pi \mathrm{G}_3} \int \mathrm{d}^3 x \sqrt{-h}\left[\frac{2}{L_3^2}+R+\ell^2\left(\frac{3}{8} R^2-R_{a b} R^{a b}\right)+\cdots\right] \\&+ \int \mathrm{d}^3 x \sqrt{-h}\left(-\frac{1}{4g_3^2}\widetilde{\mathbf{F}}^2+\frac{4 \ell_4^2}{5 g_3^2}\bar{f}\right).
\end{aligned}
\end{equation}
Furthermore, it is evident that we can define the three-dimensional cosmological constant $\Lambda_3$ in the effective theory on the brane through the AdS${}_3$ scale $L_3$ by 
\begin{equation}\label{doijepow298}
\Lambda_3=-\frac{1}{L_3^2}\,.
\end{equation}
For comparison, note that in the action \eqref{iurj389}, we can define the four-dimensional cosmological constant as
\begin{equation}\label{ej3ji928}
\Lambda_4=-\frac{3}{\ell_4^2}\,,
\end{equation}
where $\ell_4$ has the same meaning as  $L_4$ in \cite{Frassino:2022zaz}.

Using to \eqref{jeiuj3e389}, we can obtain the explicit form of  the holographic stress-energy tensor $\left\langle \bar{T}_{ab}\right\rangle$ in \eqref{dkljp9jidk83} as  
\begin{equation}\label{fejoi29}
\begin{aligned}
 8 \pi \mathrm{G}_3\left\langle \bar{T}_{a b}\right\rangle&=R_{a b}-\frac{1}{2} h_{a b}\left(R+\frac{2}{L_3^2}\right)+E_{ab}
 \\&\quad+\ell^2\left(
 \frac{1}{2} h_{a b} R_{c d} R^{c d}+\frac{3}{4} R_{a b} R-\frac{3}{16} h_{a b} R^2-2 R^{c d} R_{a c b d}\right.\\&\quad\quad\quad\quad\left. +\frac{1}{4} \nabla_b \nabla_a R-\nabla_c \nabla^c R_{a b}+\frac{1}{4} h_{a b} \nabla_c \nabla^c R\right)+\cdots\,,
 \end{aligned}
\end{equation}
where the electromagnetic term $E_{ab}$ is given by
\begin{equation}\label{jf938p2}
    E_{ab}=\frac{2 \pi  \mathrm{G}_3}{g_3^2}\left(-4 \widetilde{F}_a{ }^c \widetilde{F}_{b c}+\widetilde{\mathbf{F}}^2 g_{a b}\right)+\mathcal{O}(\ell^2)\,,
\end{equation}
where the terms at the order of $\ell^2$ is not shown as it is lengthy, though it will be used in what follows.  Then we have the trace of the electromagnetic term as
\begin{equation}\label{jf392390z}
    E^a{}_a=\frac{16 \pi  \Delta ^4 G_3 \mathcal{P}_1 q^2 \ell ^2}{2205 g_3^2 r^8 \ell _3^2 \ell _s^2}\,,
\end{equation}
where
\begin{equation}
    \mathcal{P}_1=-119357 F r \ell  \ell _3^2 \ell _4^2-801656 \mathcal{G}_3 M r^2 \ell _3^2 \ell _4^2+125424 \Delta ^4 q^2 \ell ^2 \ell _3^2 \ell _4^2+r^4 \left(2205 \ell _3^2+22658 \ell _4^2\right).
\end{equation}
To obtain \eqref{jf392390z}, we can thus obtain  the specific expression of the  holographic stress-energy tensor as
\begin{equation}\label{eji132981}
\begin{aligned}
8\pi \mathrm{G}_3\left\langle \bar{T}^b{ }_a\right\rangle &=\frac{L_3^2-\ell _3^2}{L_3^2 \ell _3^2}\mathbb{1}+\frac{F \ell }{2 \bar{r}^3}\mathrm{diag}\left(1,\,1,\,-2\right)+\mathcal{O}\left(\ell^2\right)\\&=\frac{F \ell }{2 \bar{r}^3}\mathrm{diag}\left(1,\,1,\,-2\right)+\mathcal{O}\left(\ell^2\right)\,,
\end{aligned}
\end{equation}
where in the second line \eqref{jeo3i28} has been used. The trace of the stress-energy tensor  is
\begin{equation}\label{fj2903}
    8\pi \mathrm{G}_3\left\langle \bar{T}^a{ }_a\right\rangle=\frac{\ell ^2 \mathcal{P}_2}{8820 g_3^2 \bar{r}^6 \ell _3^4 \ell_{\star}^2 \left(8 \mathcal{G}_3 M \ell _3^2-\bar{r}^2\right)}+\mathcal{O}\left(\ell^3\right)\,,
\end{equation}
where
\begin{equation}
\begin{aligned}
    \mathcal{P}_2=&1128960 \pi  \Delta ^4 \mathrm{G}_3 \mathcal{G}_3 M q^2 \bar{r}^2 \ell _3^6-141120 \pi  \Delta ^4 \mathrm{G}_3 q^2 \bar{r}^4 \ell _3^4\\&+62906880 \pi  \Delta ^4 \mathrm{G}_3 \mathcal{G}_3 M q^2 \bar{r}^2 \ell _3^4 \ell _4^2-1450112 \pi  \Delta ^4 \mathrm{G}_3 q^2 \bar{r}^4 \ell _3^2 \ell _4^2\\&-288855 g_3^2 \bar{r}^8 \ell_{\star}^2-410447872 \pi  \Delta ^4 \mathrm{G}_3 \mathcal{G}_3^2 M^2 q^2 \ell _3^6 \ell _4^2\\&+1746360 g_3^2 \mathcal{G}_3 M \bar{r}^6 \ell _3^2 \ell_{\star}^2-8820 \Delta ^4 g_3^2 q^2 \bar{r}^4 \ell _3^4 \ell_{\star}^2\\&+70560 \Delta ^4 g_3^2 \mathcal{G}_3 M q^2 \bar{r}^2 \ell _3^6 \ell_{\star}^2-4515840 g_3^2 \mathcal{G}_3^2 M^2 \bar{r}^4 \ell _3^4 \ell_{\star}^2\,.
\end{aligned}
\end{equation}
Applying \eqref{eji132981} and \eqref{fj2903}, we can see that the nonlinear Maxwell field on the brane affects the stress-energy tensor only at the order of $\ell^2$. In the bulk, we have $\mathbf{\nabla}\cdot \mathbf{F}=0$, which means that there is no current; however, it is not the same case on the brane. The induced current density on the brane is related with the stress-energy tensor of the nonlinear Maxwell field $\left\langle T_E^{a b}\right\rangle$ by
\begin{equation}
\begin{aligned}
    \left\langle J^b\right\rangle &=-\nabla_a \left\langle T_E^{a b}\right\rangle=\nabla_a\left(\frac{2}{\sqrt{-h}}\frac{\delta(\sqrt{-h} f)}{\delta \widetilde{F}_{a b}}\right)=-\frac{1}{\sqrt{-h}}\frac{\delta(\sqrt{-h} f)}{\delta \bar{\mathcal{A}}_b}\\&=-\frac{1}{g_3^2}\nabla_a F^{a b}+\frac{\ell_4^2}{g_3^2}\left(\frac{R \nabla_a \widetilde{F}^{a b}}{90}-\frac{89 \widetilde{F}^{a b} \nabla_{a}R}{90}+\frac{12}{245} \nabla_c \nabla^b \nabla_a \widetilde{F}^{a c}\right.\\&\quad\left.-\frac{562}{735} \nabla_c \nabla^c \nabla_a \widetilde{F}^{a b}+\frac{12}{245} \nabla_c \nabla^c \nabla_a \widetilde{F}^{b a}-2 R_{a c} \nabla_c \widetilde{F}^{a b}\right)\,.
\end{aligned}
\end{equation}
Explictly, we obtain
\begin{equation}
   \left\langle J^{\bar{t}}\right\rangle=-\frac{2 \Delta ^2 q \ell }{g_3^2 \bar{r}^3 \ell_{\star}}+\frac{2 \Delta ^2 q \ell ^3 \left(14352 \mathcal{G}_3 M \ell _3^2+1139 \bar{r}^2\right)}{245 g_3^2 \bar{r}^5 \ell _3^2 \ell_{\star}}+\mathcal{O}\left(\ell^4\right)\,,\quad \left\langle J^{\bar{r}}\right\rangle=0\,,\quad \left\langle J^{\bar{\phi}}\right\rangle=0\,,
\end{equation}
which is consistent with the result in \cite{Climent:2024nuj}.

We note that from \eqref{jf392390z} it is obvious that unlike the Maxwell field in the bulk, the nonlinear Maxwell field on the brane does not have the property of conformal symmetry due to the backreaction of quantum matter from the $\mathrm{CFT}_3$ on the brane. Moreover, the deviation of the nonlinear Maxwell field from  conformal symmetry is triggered by the backreaction strength parameter $\ell$ at  the order of $\mathcal{O}(\ell^2)$.

The preceding  analysis shows that the quantum charged black  hole \eqref{iroejreop1093} together with a nonlinear Maxwell field \eqref{djkoeje43} is induced on the brane. This can be viewed as an exact solution to the semiclassical equation of motion \eqref{dkljp9jidk83}  on the AdS${}_3$ brane. This quantum black hole incorporates the backreaction from the CFT\(_3\) on the brane at every order of the backreaction parameter \(\ell\). It serves as a localized gravity solution on the brane derived from the AdS charged C-metric \eqref{j9qj398} solution to the classical equation of motion \eqref{rejop3988} in the bulk.

\subsection{Thermodynamic quantities}

As suggested by \cite{Emparan:1999wa,Emparan:1999fd,Emparan:2020znc}, we can define dimensionless variables related with $\ell,\,\ell_3,\,r_+$, and $x_1$ as
\begin{equation}\label{djoieqj289}
z=\frac{\ell_3}{r_{+} x_1}\,,\quad \nu=\frac{\ell}{\ell_3}\,.
\end{equation}
Additionally, we can redefine the charge parameter as
\begin{equation}\label{2189j9pw2}
\chi=q x_1^2\,,
\end{equation}
which is dimensionless.
Then by combining $\bar{H}(\bar{r})=0$ with
\eqref{io3qjrf843} we obtain   the identities 
\begin{equation}\label{ide1}
x_1^2= - \frac{\nu ^2 \chi ^2 z^4+\nu z^3 \left(\chi ^2 -1\right)+1}{\kappa z^2 (1 +   \nu  z)}\,,
\end{equation}
\begin{equation}\label{ide2}
r_{+}^2=-\frac{\ell_3^2 \kappa  (1 +   \nu  z)}{\nu ^2 \chi ^2 z^4+\nu  \left(\chi ^2-1\right) z^3+1}\,,
\end{equation}
\begin{equation}\label{ide3}
\mu x_1=-\frac{\kappa  \left(\nu ^2 \chi ^2 z^4-\left(\chi ^2-1\right) z^2+1\right)}{\nu ^2 \chi ^2 z^4+\nu  \left(\chi ^2-1\right) z^3+1}\,.
\end{equation}
Since $x_1^2,\,r_+^2,$, and $\mu x_1$ cannot be negative,  the parameter $\kappa$ defined in \eqref{dkjaoe834} is now determined by 
\begin{equation}
\kappa=-\operatorname{sgn}\left(\nu ^2 \chi ^2 z^4+\nu  \chi ^2 z^3-\nu  z^3+1\right)\,,
\end{equation}
where $\operatorname{sgn}$ is the sign function.

As a result, $F(M)$ in \eqref{398ur82jemd} can be rewritten as 
\begin{equation}
F(M)=\frac{8 z^4 (\nu  z+1)^2 \left(\nu ^2 \chi ^2 z^4-\left(\chi ^2-1\right) z^2+1\right)}{\left(\nu ^2 \chi ^2 z^4+2 \nu  \left(\chi ^2+1\right) z^3+\left(\chi ^2+3\right) z^2+1\right)^3}\,,
\end{equation}
where the identities \eqref{ide1}-\eqref{ide2} are used. The mass of the three-dimensional quantum charged black hole, which is related to the conical deficit given by \eqref{joepj398}, can now be rewritten as
\begin{equation}\label{dkje842}
\begin{aligned}
M&=-\frac{\ell}{2 \mathrm{G}_3\ell_4}  \frac{\kappa x_1^2}{\left(3-\kappa x_1^2+q^2 x_1^4\right)^2}\\&=\frac{z^2 (\nu  z+1) \left(\nu ^2 \chi ^2 z^4+\nu  \left(\chi ^2-1\right) z^3+1\right)}{2 \mathcal{G}_3 \left(\nu ^2 \chi ^2 z^4+2 \nu  \left(\chi ^2+1\right) z^3+\left(\chi ^2+3\right) z^2+1\right)^2}\,,
\end{aligned}
\end{equation}
where in the last step we have used the identities \eqref{dejoiefjre}, which can be further written as
\begin{equation}\label{rioj4820}
\mathcal{G}_3=\frac{\mathrm{G}_3}{\sqrt{1+\nu^2}}\,.
\end{equation}

To calculate the entropy of the quantum charged black hole, we first compute the entropy associated with the bulk horizon using the bulk metric given by  \eqref{j9qj398}. The result is 
\begin{equation}\label{fjeoir3}
\begin{aligned}
S_{\text {gen }}&=\frac{2}{4 \mathrm{G}_4} \int_0^{2 \pi \Delta} \mathrm{d} \phi \int_0^{x_1} \mathrm{d} x r_{+}^2 \frac{\ell^2}{\left(\ell+r_{+} x\right)^2}\\&=\frac{\pi  \ell_3 \sqrt{\nu ^2+1} z}{\mathrm{G}_3 \left(\nu ^2 \chi ^2 z^4+2 \nu  \left(\chi ^2+1\right) z^3+\left(\chi ^2+3\right) z^2+1\right)}\\&=\frac{\pi  \ell_3 z}{\mathcal{G}_3 \left(\nu ^2 \chi ^2 z^4+2 \nu  \left(\chi ^2+1\right) z^3+\left(\chi ^2+3\right) z^2+1\right)}\,,
\end{aligned}
\end{equation}
where we denote the bulk entropy as the general entropy $S_{\text {gen }}$ \cite{Emparan:2020znc} consisting of the Wald entropy \cite{Wald:1993nt} and the entanglement entropy \cite{Emparan:2006ni} and the factor of 2 originates from the left-right symmetric configuration of the quantum black hole with two branes glued together. We can view this entropy as a generalized entropy that incorporates the  all orders of backreaction of the quantum matter for the quantum black hole. 

For comparison, we note that there can be two other kinds of entropies for the quantum black hole, i.e., the Bekenstein-Hawking entropy and the Wald entropy. Employing \eqref{iroejreop1093}, the Bekenstein-Hawking area entropy of the quantum charged black hole on the brane is
\begin{equation}
S_{\mathrm{cl}}=\frac{2 \pi r_{+} \Delta}{4 \mathrm{G}_3} =\frac{1+\nu z}{\sqrt{1+\nu^2}} S_{\text {gen }}=\left(1-\frac{\nu ^2}{2}-\frac{\nu ^3 z}{2}+\nu  z+\mathcal{O}\left(\nu^4\right)\right)S_{\text {gen }}\,.
\end{equation}
This area entropy is, however, principally not suitable for the quantum charged black hole on the brane. The reason is that  the gravity on the brane is a higher curvature theory, and for the quantum charged black hole, there is also a nonlinear electromagnetic field coupled with the background spacetime curvature. Furthermore, as pointed out in \cite{Emparan:2020znc} for the quBTZ black hole, the three-dimensional area entropy  $S_{\mathrm{cl}}$ cannot satisfy the thermodynamic first law of the black hole as zero and infinite derivatives emerge when differentiating it with respect to mass. 

The Wald entropy \cite{Wald:1993nt} $S_W$ in such a sense thus can be an effective candidate entropy for this gravitational theory, which, according to the formula presented in \cite{Jacobson:1993vj}, is (see appendix \ref{owi23i8} for details)
\begin{equation}\label{wjio29p3}
    S_W=\frac{\pi  z \ell _3 \left(\nu ^3 \chi ^2 z^3 (\nu  z+1) (3 \nu  z+2)-\nu  \left(\nu +\nu ^2 z \left(2 z^2+3\right)-2 z\right)+2\right)}{2 \mathcal{G}_3 \sqrt{\nu ^2+1} \left(z^2 \left((\chi +\nu  \chi  z)^2+2 \nu  z+3\right)+1\right)}\,.
\end{equation}
Note that the Wald entropy given by \eqref{wjio29p3} is valid only in the small $\nu$ limit up to $\mathcal{O}(\nu^4)$. However, we aim to obtain an entropy that accounts for the backreaction of the quantum corrections from CFT on the brane at every order of $\nu$. As a result, the generalized entropy $S_{\text {gen }}$ \eqref{fjeoir3} for the classical four-dimensional black hole can be the only proper candidate mapping to the three-dimensional quantum charged black hole. As we will see, this entropy indeed satisfies the first law of the quantum charged black hole. We can also obtain the four-dimensional area entropy $S_{\text {gen }}$ as the entropy of the quantum black hole by the Euclidean method, just as has been checked by \cite{Kudoh:2004ub} for the quBTZ black hole. When the backreaction from the CFT${}_3$ vanishes ($\nu\to 0$), the differences between the generalized entropy, the area entropy, and the Wald entropy vanish, yielding
\begin{equation}\label{jfep9i3p28}
\left.S_{\text {gen }}\right|_{\nu=0}=\left.S_{\mathrm{cl}}\right|_{\nu=0}=\left.S_{W}\right|_{\nu=0}=\frac{\pi  \ell_3 z}{\mathrm{G}_3\left( \left(\chi ^2+3\right) z^2+1\right)}\,.
\end{equation}
The canonical timelike Killing vector of the quantum charged black hole on the brane is $\partial / \partial \bar{t}$, which gives the temperature of the black hole on the quantum horizon as
\begin{equation}
\begin{aligned}
 T=\frac{\Delta H^{\prime}\left(r_{+}\right)}{4 \pi} =\frac{z \left(\nu  z \left(z^2 \left(1-(\chi +\nu  \chi  z)^2\right)+3\right)+2\right)}{2 \pi  \ell_3 \left(z^2 \left((\chi +\nu  \chi  z)^2+2 \nu  z+3\right)+1\right)}\,.
\end{aligned}
\end{equation}

To calculate the electric charge of the  quantum black hole, we identify the charge computed on the brane with that of the bulk. Using the gauge potential \eqref{jsp39} together with the action \eqref{j201dm} and the relation \eqref{j2091}, we obtain for the latter
\begin{equation}\label{ejiop3i4j}
\begin{aligned}
    Q&=\frac{1}{g_4^2}  \int * \mathrm{\mathbf{F}}\\&=\frac{4q\ell}{g_4^2 \ell_{\star}}\int_0^{2 \pi\Delta} \mathrm{d} \phi \int_0^{x_1} \mathrm{d} x\\&=\frac{2 \sqrt{2 \pi  \nu  \ell _3}}{g_4 \sqrt{\mathcal{G}_3}}\frac{\chi  z^2 (\nu  z+1)}{ z^2 \left((\chi +\nu  \chi  z)^2+2 \nu  z+3\right)+1}\,.
\end{aligned}
\end{equation}
We also find 
\begin{equation}\label{jofiej932}
\Phi=\left.\bar{\mathcal{A}}_\mu \xi^{\mu}\right|_{r \rightarrow r_{+}} ^{r \rightarrow \infty}=g_4 \sqrt{\frac{\nu }{2 \pi  \mathcal{G}_3 \ell _3}}\frac{\chi  z^3 (\nu  z+1)}{\nu ^2 \chi ^2 z^4+2 \nu  \left(\chi ^2+1\right) z^3+\left(\chi ^2+3\right) z^2+1}
\end{equation}
 for  the conjugate  electric potential on the brane, 
where we choose the Killing vector $\xi^a$ to be $\partial/\partial\bar{t}$, and we have used the gauge potential \eqref{djkoeje43}. An alternative way to obtain \eqref{jofiej932}  is to use the gauge potential   \eqref{jsp39} for the bulk spacetime with the canonical time coordinate $\bar{t}$ in \eqref{jeioe3}.

According to the above results, it is straightforward to verify that the  relations
\begin{equation}
\partial_z M-T \partial_z S_{\text {gen }}-\Phi \partial_z Q=0\,,
\end{equation}
\begin{equation}
\partial_\chi M-T \partial_\chi S_{\text {gen }}-\Phi \partial_\chi Q=0
\end{equation}
are satisfied, where we express thermodynamic quantities in terms of the `renormalized' Newton constant $\mathcal{G}_3$. These two relations ensure that the thermodynamic first law of the quantum charged black hole is 
\begin{equation}
\mathrm{d} M-T \mathrm{d} S_{\mathrm{gen}}-\Phi \mathrm{d} Q=0\,.
\end{equation}
This universal differential relation applies for all parameters for the thermodynamic quantities of the quantum charged black hole.

\section{Thermodynamics of quantum charged black holes}\label{dfjj39882}

We will now investigate the thermodynamic first law and Smarr (energy) relation \cite{Smarr:1972kt,Kastor:2009wy} in the extended phase space of the quantum charged black hole, where the cosmological constant is viewed as a dynamical quantity \cite{Kubiznak:2016qmn}. We shall examine this from three perspectives -- bulk, brane, and boundary -- in the holographic braneworld model.

\subsection{Bulk description}

\begin{figure}
    \centering
    \includegraphics[width=0.4\linewidth]{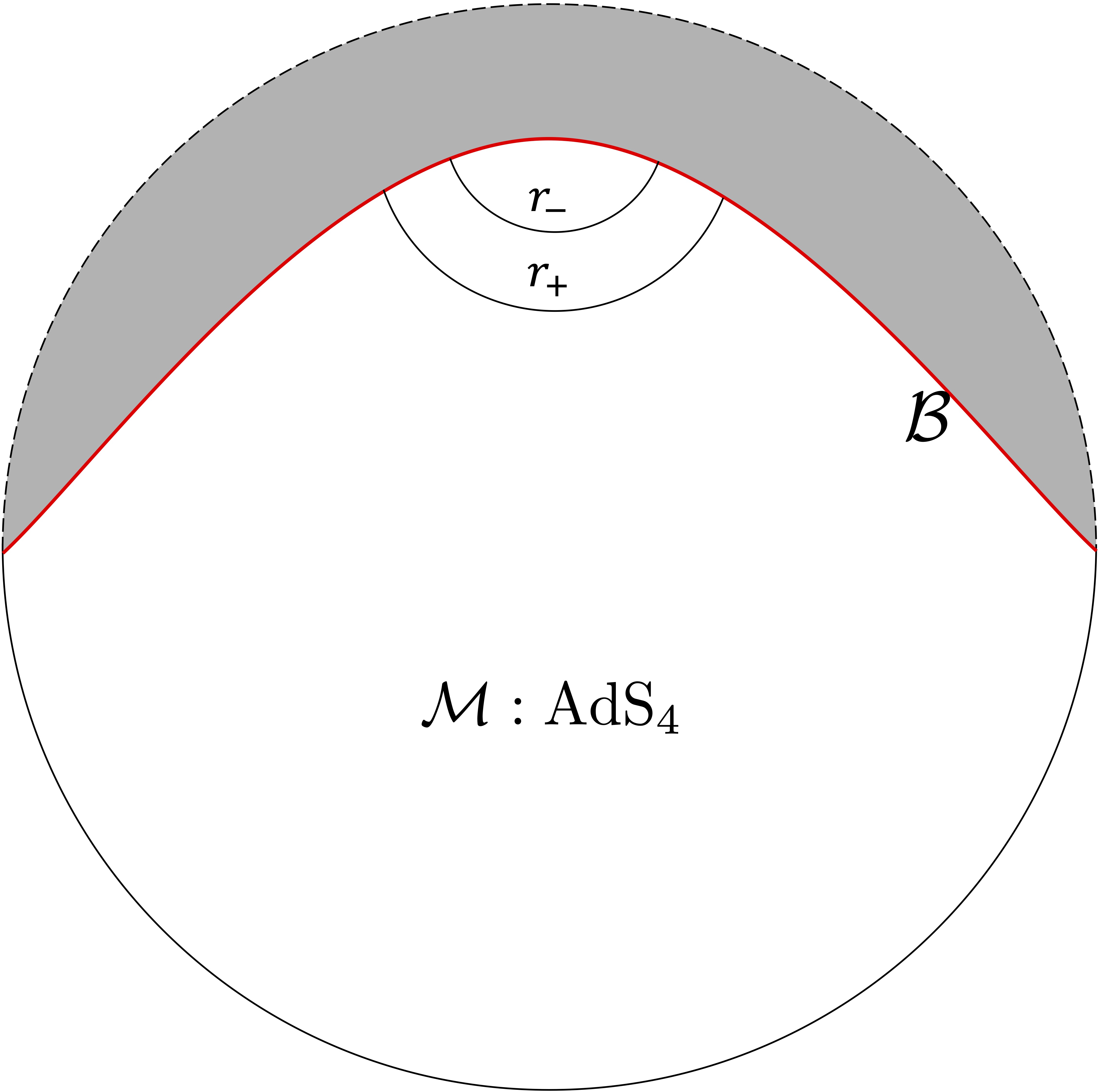}
    \caption{Sketch of the bulk perspective. The bulk is the region enclosed by the three-dimensional KR brane (red curve) and the half-space of the asymptotic boundary (solid black curve). From the perspective of the bulk, the thermodynamic system consists of the classical black hole with inner and outer horizons $r_\mp$ in the $\mathrm{AdS}_4$ bulk $\mathcal{M}$ and a brane $\mathcal{B}$ with a variable tension.}
    \label{ji3489j}
\end{figure}

Let's first consider the thermodynamics of the quantum charged black hole from the bulk perspective. As shown in Fig. \ref{ji3489j}, the thermodynamic system consists of the bulk black hole and a KR brane with variable tension. From the pure bulk perspective of this holographic braneworld setup, we have an $\mathrm{AdS}_{4}$ bulk spacetime $\mathcal{M}$, coupled with an $\mathrm{AdS}_{3}$ brane $\mathcal{B}$ with a tension $\tau$. The quantum charged black hole on the brane corresponds to the classical solution of the bulk gravity with a brane and encodes all orders of backreactions from the $\mathrm{CFT}_3$ residing on the brane. In the extended thermodynamic phase space, we can view the four-dimensional cosmological constant $\Lambda_4$ and the brane tension $\tau$ as variables. The variation of $\Lambda_4$ yields the four-dimensional pressure $P_4$ and its conjugate four-dimensional volume $V_4$. The variation of $\tau$ amounts to altering the position of the brane according to the Israel junction conditions. The quantity conjugate to $\tau$ has a dimension of area; we shall refer to it as the thermodynamic area of the brane, denoting it as $A_\tau$, so that $A_\tau \mathrm{d}\tau$ is a work term.

According to \eqref{eajj38}, \eqref{dkje842}, \eqref{fjeoir3}, and \eqref{ejiop3i4j}, the mass $M$, entropy $S$, and electric charge $Q$ of the quantum charged black hole from the bulk viewpoint are expressed in terms of the four-dimensional Newton constant $\mathrm{G}_4$ as
\begin{eqnarray}
    M&=&\frac{z^2 \ell  (\nu  z+1) \left(\nu ^2 \chi ^2 z^4+\nu  \left(\chi ^2-1\right) z^3+1\right)}{\mathrm{G}_4 \left(\nu ^2 \chi ^2 z^4+2 \nu  \left(\chi ^2+1\right) z^3+\left(\chi ^2+3\right) z^2+1\right)^2}\,,\\
    S&=&S_{\mathrm{gen}}=\frac{2 \pi  z \ell ^2}{\mathrm{G}_4 \nu  \left(\nu ^2 \chi ^2 z^4+2 \nu  \left(\chi ^2+1\right) z^3+\left(\chi ^2+3\right) z^2+1\right)}\,,\\
    Q&=&\frac{4 \sqrt{\pi } \chi  z^2 \ell  (\nu  z+1)}{\sqrt{\mathrm{G}_4} \left(g_4 z^2 \left((\chi +\nu  \chi  z)^2+2 \nu  z+3\right)+g_4\right)}\,.
\end{eqnarray}
Furthermore, the thermodynamic pressure of the bulk spacetime is \cite{Kubiznak:2016qmn}
\begin{equation}
P_4=-\frac{\Lambda_4}{8\pi \mathrm{G}_4}=\frac{3 \left(\nu ^2+1\right)}{8 \pi  \mathrm{G}_4 \ell ^2}\,,
\end{equation}
where in the last step we have used \eqref{dklaje292}, \eqref{ej3ji928}, and \eqref{djoieqj289}. As noted above, the tension \eqref{fjioe3894} of the brane can also be viewed as a thermodynamic variable, which is given in terms of $\mathrm{G}_4$ and the backreaction parameter $\ell$. This backreaction parameter $\ell$ also appears in the expressions for $M,\,S\,,Q,$ and $P_4$.

According to the above setup, we can calculate the temperature $T$, electric potential $\Phi$, four-dimensional thermodynamic volume $V_4$, and thermodynamic area $A_\tau$ for the brane as
\begin{eqnarray}\label{eipo2981}
T &=&\left(\frac{\partial M}{\partial S}\right)_{Q,\, P_4,\, \tau}=-\frac{\nu  z \left(\nu ^3 \chi ^2 z^5+2 \nu ^2 \chi ^2 z^4+\nu  \left(\chi ^2-1\right) z^3-3 \nu  z-2\right)}{2 \pi  \ell \left(\nu ^2 \chi ^2 z^4+2 \nu  \left(\chi ^2+1\right) z^3+\left(\chi ^2+3\right) z^2+1\right)}\,,\label{j929j23}
\\
\label{jkaeir9328}
\Phi &=&\left(\frac{\partial M}{\partial Q}\right)_{S, \,P_4,\, \tau}=\frac{g_4 \nu  \chi  z^3 (\nu  z+1)}{\sqrt{\pi \mathrm{G}_4} \left(z^2 \left((\chi +\nu  \chi  z)^2+2 \nu  z+3\right)+1\right)}\,,\label{2j9pi1}
\\
V_4 &=& \left(\frac{\partial M}{\partial P_4}\right)_{S, \,Q,\, \tau}=\frac{8 \pi  z^2 \ell ^3 (2 \nu  z+1)}{3 \nu ^2 \left(\nu ^2 \chi ^2 z^4+2 \nu  \left(\chi ^2+1\right) z^3+\left(\chi ^2+3\right) z^2+1\right)^2}\,,
\\
A_\tau &=& \left(\frac{\partial M}{\partial \tau}\right)_{S, \,Q,\, P_4} \nonumber\\
&=& \frac{2 \pi  z^2 \ell ^2 (\nu  z+1) \left(\nu ^2+\nu ^4 \chi ^2 z^4+\nu ^3 \left(\chi ^2+1\right) z^3+2 \nu ^2 z^2-2 \nu  z-2\right)}{\nu ^2 \left(\nu ^2 \chi ^2 z^4+2 \nu  \left(\chi ^2+1\right) z^3+\left(\chi ^2+3\right) z^2+1\right)^2}\,.
\end{eqnarray}
Here, we have kept the four-dimensional Newton constant $\mathrm{G}_4$ fixed, which is a natural setting. The electric potential \eqref{jkaeir9328} we obtained here is identical to the one in \eqref{jofiej932}. From the bulk perspective, we find that the thermodynamic first law in the extended phase space, where the four-dimensional cosmological constant $\Lambda_4$ and the brane tension $\tau$ are viewed as thermodynamic variables, can be written as
\begin{equation}
\mathrm{d}M=T \mathrm{d} S+\Phi \mathrm{d}Q+V_4 \mathrm{d}P_4+A_\tau \mathrm{d}\tau 
\end{equation}
and the Smarr relation
\begin{equation}
M=2TS-2V_4P_4- A_\tau \tau +\Phi Q
\end{equation}
are satisfied. Note that this latter relation can be obtained by Euler’s theorem \cite{Frassino:2022zaz}, as $\sqrt{G_4}M$ is a homogeneous function of $\mathrm{G}_4 S,\,\Lambda_4,\,\mathrm{G}_4\tau,\,$ and $\sqrt{\mathrm{G}_4} Q$.

\subsection{Brane description}

\begin{figure}
    \centering
    \includegraphics[width=0.4\linewidth]{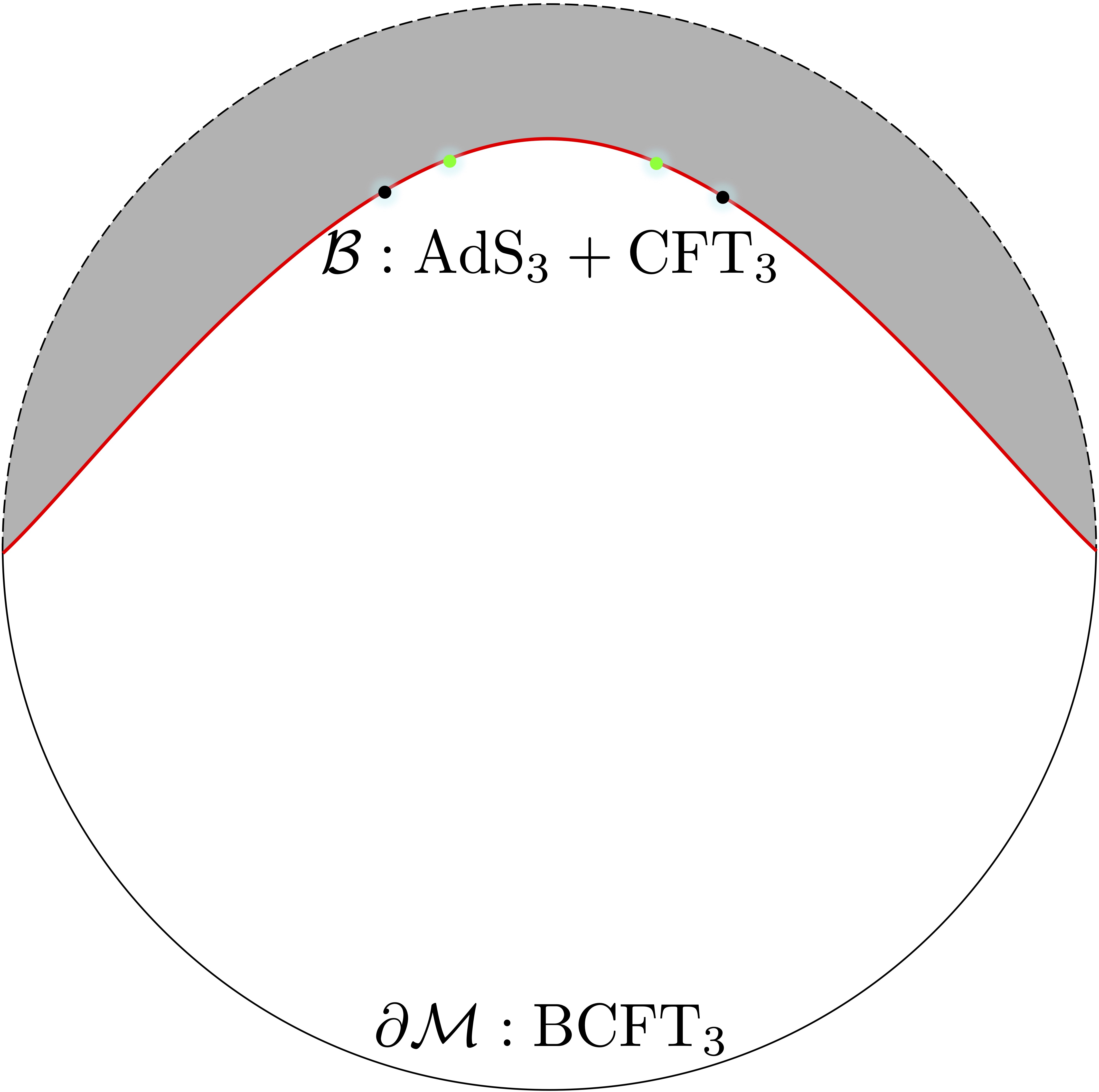}
    \caption{Sketch of the brane perspective. The thermodynamics of the quantum charged black hole dwelling on the AdS brane $\mathcal{B}$ (red curve) with inner and outer horizons $r_\mp$ (green and black dots) receiving quantum backreaction from the coupled holographic $\mathrm{CFT}_3$ on the brane, which communicates with the $\mathrm{BCFT}_3$ at the asymptotic boundary $\partial\mathcal{M}$.}
    \label{ji3ejop3pqo}
\end{figure}
Next, we consider the thermodynamics of the quantum charged black hole from the perspective of the $\mathrm{AdS}_3$ brane (see Fig. \ref{ji3ejop3pqo}). In the holographic KR braneworld, the $\mathrm{AdS}_3$ brane intersects the asymptotic boundary of the bulk spacetime at two defects. The brane serves as a holographic renormalization surface for the asymptotic $\mathrm{CFT}_3$, and the local higher curvature gravitational theory on the brane receives backreaction effects from the $\mathrm{CFT}_3$ on the brane. The UV microscopic degrees of freedom for the $\mathrm{CFT}_3$ are removed by the brane, leaving its central charge to be \cite{Emparan:2020znc,Emparan:2021hyr}
\begin{equation}\label{jfeorj39822}
c_3=\frac{\ell_4^2}{\mathrm{G}_4}=\frac{\ell}{2 \mathrm{G}_3 \sqrt{\nu ^2+1}}=\frac{\nu  \ell _3}{2 \mathrm{G}_3 \sqrt{\nu ^2+1}}\,,
\end{equation}
where in the last step we have used \eqref{dklaje292} and \eqref{dejoiefjre} to express the central charge for the $\mathrm{CFT}_3$ on the brane in terms of the $\mathrm{AdS}_3$ radius $\ell_3$ and the three-dimensional Newton constant $\mathrm{G}_3$. From \eqref{jfeorj39822}, we also note that the central charge of the $\mathrm{CFT}_3$ is affected by the backreaction parameter $\ell$. This suggests that the central charge $c_3$ can be regarded as a thermodynamic variable \cite{Cong:2021fnf,Cong:2021jgb}. Different from the method used in \cite{Cong:2021fnf,Cong:2021jgb}, we here will keep the three-dimensional Newton constant $\mathrm{G}_3$ fixed\footnote{See \cite{Susskind:2021nqs} for a discussion of a running Newton constant in the context of string/black hole transitions \cite{Horowitz:1997jc,Chen:2021dsw,Balthazar:2022hno,Ceplak:2023afb}.}. Our approach is similar to that employed in \cite{Karch:2015rpa,Sinamuli:2017rhp,Visser:2021eqk,Ahmed:2023snm,Ahmed:2023dnh,Zhang:2023uay,Gong:2023ywu}; see \cite{Mann:2024sru} for a recent review on this topic.

In the extended phase space approach, the thermodynamic pressure corresponds to the three-dimensional cosmological constant $\Lambda_3$ in \eqref{doijepow298}. Using \eqref{dklaje292}, \eqref{jeo3i28}, and \eqref{djoieqj289}, we obtain
\begin{equation}
P_3=-\frac{\Lambda_3}{8\pi \mathrm{G}_3}=\frac{\sqrt{\nu ^2+1}}{4 \pi  \mathrm{G}_3 \left(\sqrt{\nu ^2+1}+1\right) \ell _3^2}\,, 
\end{equation}
which simplifies to $P_3=1/\left(8\pi \mathrm{G}_3 \ell_3^2\right)$ when the strength of the backreaction goes to zero. Reexpressing the mass \eqref{dkje842}, entropy \eqref{fjeoir3}, and electric charge \eqref{ejiop3i4j} for the quantum charged black hole on the brane, we get
\begin{eqnarray}
\label{dji329p}
M &=& \frac{\sqrt{\nu ^2+1} z^2 (\nu  z+1) \left(\nu ^2 \chi ^2 z^4+\nu  \left(\chi ^2-1\right) z^3+1\right)}{2 \mathrm{G}_3 \left(\nu ^2 \chi ^2 z^4+2 \nu  \left(\chi ^2+1\right) z^3+\left(\chi ^2+3\right) z^2+1\right)^2}\,,
\\ 
S_{\mathrm{gen}} &=& \frac{\pi  \ell_3 \sqrt{\nu ^2+1} z}{\mathrm{G}_3 \left(\nu ^2 \chi ^2 z^4+2 \nu  \left(\chi ^2+1\right) z^3+\left(\chi ^2+3\right) z^2+1\right)}\,,
\\ 
Q &=& \sqrt{\frac{\pi  \left(\nu ^2+1\right)}{5 \mathrm{G}_3}}\frac{4 \chi  z^2 (\nu  z+1)}{g_3 z^2 \left((\chi +\nu  \chi  z)^2+2 \nu  z+3\right)+g_3}
\end{eqnarray}
in terms of the three-dimensional Newton constant $\mathrm{G}_3$. The thermodynamics of the quantum charged black hole encodes the backreaction of the quantum matter on the $\mathrm{CFT}_3$.

Keeping $\mathrm{G}_3$ as a constant, the temperature $T$, electric potential $\Phi$, three-dimensional thermodynamic volume $V_3$, and three-dimensional chemical potential for the quantum charged black hole $\mu_3$ are
\begin{eqnarray}
T &=& \left(\frac{\partial M}{\partial S_{\mathrm{gen}}}\right)_{Q,\, P_3,\, c_3}=\frac{z \left(\nu  z \left(z^2 \left(1-(\chi +\nu  \chi  z)^2\right)+3\right)+2\right)}{2 \pi  \ell _3 \left(z^2 \left((\chi +\nu  \chi  z)^2+2 \nu  z+3\right)+1\right)}\,,\label{jp9i23p}
\\
\label{jkaeir932k8}
\Phi &=& \left(\frac{\partial M}{\partial Q}\right)_{S_{\mathrm{gen}},\, P_3,\, c_3}
=\sqrt{\frac{5}{4 \pi  \mathrm{G}_3}}\frac{g_3 \nu  \chi  z^3 (\nu  z+1)}{z^2 \left((\chi +\nu  \chi  z)^2+2 \nu  z+3\right)+1}\,,\label{oipe3982}
\\
V_3 &=& \left(\frac{\partial M}{\partial P_3}\right)_{S_{\mathrm{gen}},\, Q,\, c_3} 
\nonumber\\
&=& -\frac{2 \pi  z^2 \ell _3^2 (\nu  z+1) \left(\nu  \left(\nu +z \left(\nu  z \left(\nu  \chi ^2 z (\nu  z+1)+\nu  z+2\right)-2\right)\right)-2\right)}{\left(z^2 \left((\chi +\nu  \chi  z)^2+2 \nu  z+3\right)+1\right)^2}\,,
\\
\mu_3 &=& \left(\frac{\partial M}{\partial c_3}\right)_{S_{\mathrm{gen}},\, Q, \,P_3}
\nonumber\\
&=&\frac{\left(\chi ^2-1\right) z^5+z^3}{\ell _3 \left(\left(\chi ^2+3\right) z^2+1\right)^2}\label{fj392p}
 \\
&&-\frac{\nu  z^2 \left(3 \left(\chi ^2+3\right) z^2+4 z^4 \left(\chi ^2+\left(\chi ^4-3 \chi ^2-2\right) z^2+2\right)+3\right)}{2 \ell _3 \left(\left(\chi ^2+3\right) z^2+1\right)^3}+ \mathcal{O}\left(\nu^2\right)\,, \nonumber
\end{eqnarray}
where the temperature and electric potential are the same as their respective bulk counterparts in \eqref{eipo2981} and \eqref{2j9pi1}. This is not difficult to understand, as fixing $P_4$ and $\tau$ yields $\mathrm{d}\ell=0=\mathrm{d}\nu$ in \eqref{j929j23} and \eqref{2j9pi1}, and fixing $P_3$ and $c_3$ yields $\mathrm{d}\ell_3=0=\mathrm{d}\nu$ in \eqref{jp9i23p} and \eqref{oipe3982}, which, according to \eqref{djoieqj289}, also gives $\mathrm{d}\ell=0$. We here only show the expression for $\mu_3$ to the linear order of $\nu$ in \eqref{fj392p}; see Appendix \ref{joiew} for the full expression.  From these relationships, it is straightforward to verify that the first law 
\begin{equation}\label{FLbrane}
\mathrm{d}M=T\mathrm{d}S_{\mathrm{gen}}+\Phi \mathrm{d}Q+V_3 \mathrm{d}P_3+\mu_3 \mathrm{d}c_3 
\end{equation}
and the Smarr relation
\begin{equation}\label{wekl19}
0=T S_{\text {gen }}-2 P_3 V_3+\mu_3 c_3
\end{equation}
for the quantum charged black hole are both satisfied when viewed from the perspective of the brane. The Smarr mass relation is consistent with a scaling argument as $\mathrm{G}_3M$ is a homogeneous function of $\mathrm{G}_3 S_{\mathrm{gen}},\,\Lambda_3,\,$ and $\mathrm{G}_3 c_3,\,$. The absence of $M$ and $Q$ in the Smarr relation is due to $\sqrt{\mathrm{G}_3}M$ and $Q$ being dimensionless.

The form of the first law given in \eqref{FLbrane} is illuminating, as all thermodynamic quantities involved are three-dimensional. These quantities pertain either to the black hole or to the $\mathrm{CFT}_3$ with a central charge $c_3$ and its conjugate chemical potential $\mu_3$ on the brane. This equation, in fact, represents a mixed form of the first law, comprising both brane quantities and CFT quantities, which is quite similar to the results obtained in \cite{Cong:2021fnf,Cong:2021jgb}. In this context, there may be central charge criticality for the quantum charge black hole on the brane; this is a topic we leave for future investigation. It's also worth noting that in the Smarr relation \eqref{wekl19}, the electric charge is absent since it is dimensionless in the three-dimensional brane setup. This result differs from that obtained in \cite{Frassino:2015oca} for the three-dimensional classical charged BTZ black hole, where the charge is presented in the mass formula via the introduction of a thermodynamic renormalization length scale.

\subsection{Boundary description}
\begin{figure}
    \centering
    \includegraphics[width=0.4\linewidth]{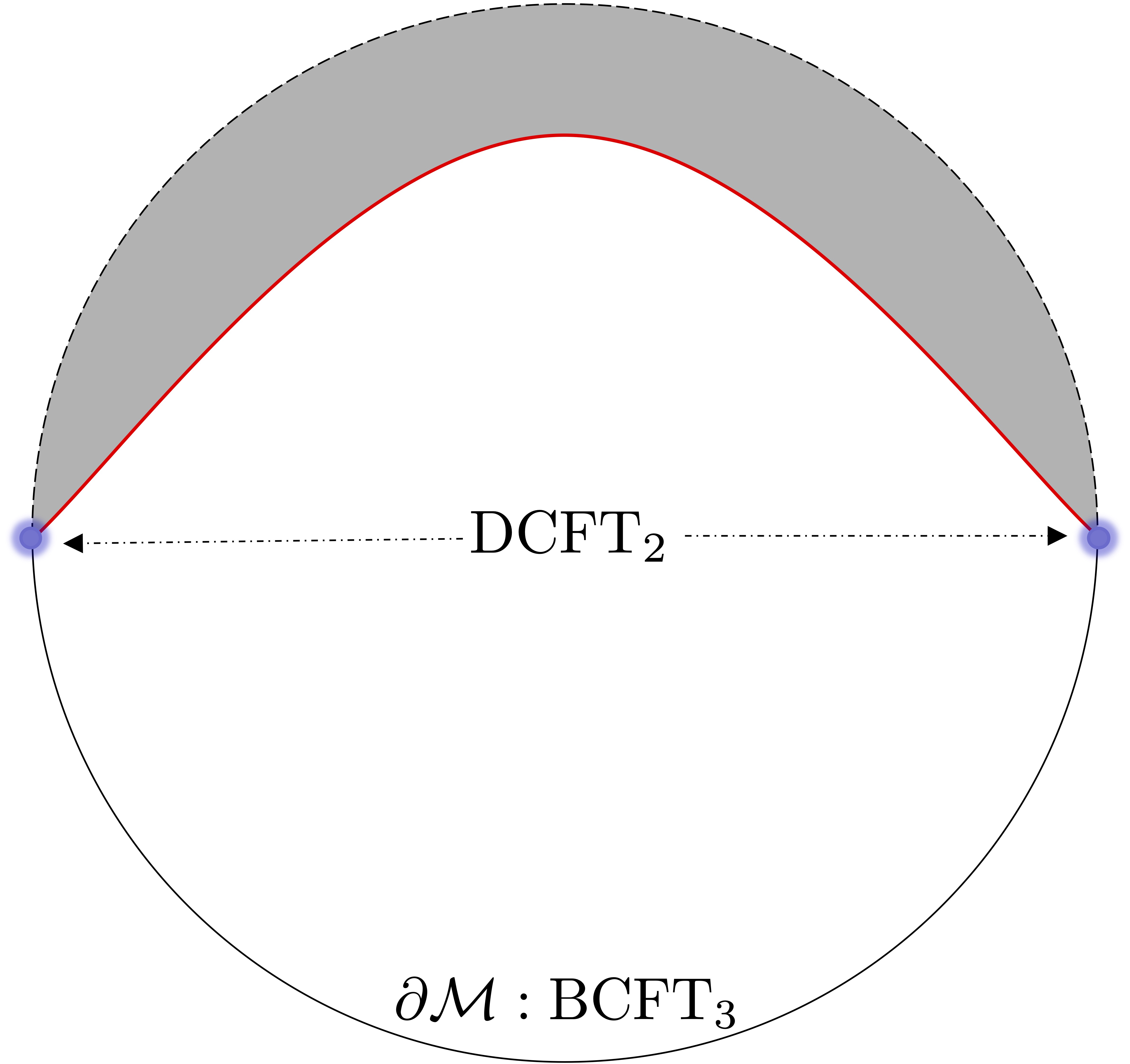}
    \caption{Sketch of the pure boundary perspective. The KR brane intersects the $\mathrm{BCFT}_3$ (black curve) at two defects (purple points). In the pure boundary perspective, we study the thermodynamics of the quantum charged black hole on the brane via the dual $\mathrm{DCFT}_2$, which is coupled with $\mathrm{BCFT}_3$ at the asymptotic boundary $\partial\mathcal{M}$.}
    \label{jpqo289}
\end{figure}

From the bulk perspective, the thermodynamic system comprises a four-dimensional bulk black hole spacetime coupled to a three-dimensional KR brane. From the brane perspective, however, the thermodynamic system transits to a quantum charged black hole spacetime coupled with a three-dimensional $\mathrm{CFT}$ on the $\mathrm{AdS}_3$ two-brane. As depicted in Fig. \ref{jpqo289}, the brane intersects the $\mathrm{BCFT}_3$, resulting in the formation of two defects, each hosting a two-dimensional  $\mathrm{DCFT}$. According to the double holography prescription, this $\mathrm{DCFT}_2$, coupled to the $\mathrm{BCFT}_3$ and in conjunction with the dynamical higher curvature gravity on the AdS${}_3$ brane, provides a dual description of the thermodynamics on the brane.

Let's consider the thermodynamics of the quantum charged black hole from the boundary perspective, i.e., from the viewpoint of the $\mathrm{DCFT}_2$, which is coupled with the $\mathrm{BCFT}_3$. The degrees of freedom for the  $\mathrm{DCFT}_2$ are related to the three-dimensional cosmological constant $L_3$ and the Newton constant $\mathrm{G}_3$ through the holographic correspondence formula \cite{Brown:1986nw}
\begin{equation}
c_2=\frac{3 L_3}{2 \mathrm{G}_3}=\frac{3 \ell }{2 \mathrm{G}_3 \sqrt{2 \nu ^2-2 \sqrt{\nu ^2+1}+2}}\,.
\end{equation}
The coefficient here is not crucial and will not affect our results qualitatively.
According to the holographic dictionary, the energy $E$, entropy $\mathcal{S}$, electric charge $\mathcal{Q}$, and volume $V_2$ for the  $\mathrm{DCFT}_2$ are respectively given by
\begin{eqnarray}
    E&=& M\,, \\
    \mathcal{S} &=& S_{\mathrm{gen}}=\frac{\pi  \sqrt{\nu ^2+1} z \ell }{\mathrm{G}_3 \nu  \left(\nu ^2 \chi ^2 z^4+2 \nu  \left(\chi ^2+1\right) z^3+\left(\chi ^2+3\right) z^2+1\right)}\,, \\
\mathcal{Q}&=&Q L_3=\sqrt{\frac{8 \pi }{5 \mathrm{G}_3}} \frac{\sqrt[4]{\nu ^2+1} \chi  z^2 \ell  (\nu  z+1)}{\sqrt{\sqrt{\nu ^2+1}-1} \left(g_3 z^2 \left((\chi +\nu  \chi  z)^2+2 \nu  z+3\right)+g_3\right)} \,,\label{j2998jd}  \\
V_2&=&2\pi L_3=\frac{2 \pi  \ell }{\sqrt{2 \nu ^2-2 \sqrt{\nu ^2+1}+2}}\,,
\label{V2eq}
\end{eqnarray}
where the specific form of the energy $E$ for the  $\mathrm{DCFT}_2$ was chosen to match \eqref{dji329p} and the electric charge $\mathcal{Q}$ on the $\mathrm{DCFT}_2$ was rescaled by $L_3$ \cite{Karch:2015rpa,Visser:2021eqk,Cong:2021jgb}. To obtain the thermodynamic volume $V_2$ for the $\mathrm{DCFT}_2$, we can write the boundary metric as \cite{Ahmed:2023snm}
\begin{equation}
    \mathrm{d} s^2_{\mathrm{bdy}}=\omega^2\left(-\mathrm{d} \bar{t}^2+L_3^2 \mathrm{d}\bar{\phi}^2\right)\,,
\end{equation}
where $\omega$ is a dynamical conformal factor, implying $V_2\propto \omega L_3$; we have set $\omega=1$ in \eqref{V2eq}.
 
Based on the above results, the temperature $T$, two-dimensional pressure $P_2$, two-dimensional chemical potential $\mu_2$, and electric potential $\phi$ for the $\mathrm{DCFT}_2$ are given by
\begin{eqnarray}
 T &=& \left(\frac{\partial E}{\partial \mathcal{S}}\right)_{\ell,\, V_2,\, c_2,\, \mathcal{Q}}=-\frac{\nu  z \left(\nu ^3 \chi ^2 z^5+2 \nu ^2 \chi ^2 z^4+\nu  \left(\chi ^2-1\right) z^3-3 \nu  z-2\right)}{2 \pi  \ell  \left(\nu ^2 \chi ^2 z^4+2 \nu  \left(\chi ^2+1\right) z^3+\left(\chi ^2+3\right) z^2+1\right)}\,, \label{9p23j92} \\
 P_2 &=& -\left(\frac{\partial E}{\partial V_2}\right)_{\ell,\, S, \,c_2, \mathcal{Q}}=\frac{\nu  z^2}{4 \pi  \mathrm{G}_3 \ell  \left(\left(\chi ^2+3\right) z^2+1\right)^2}+\mathcal{O}(\nu^2)\,, \label{2j893qdp}
 \\
 \mu_2 &=&\left(\frac{\partial E}{\partial c_2}\right)_{\ell,\, S, \,V_2, \mathcal{Q}}=\frac{ \sqrt{2\nu ^2+2} \sqrt{\nu ^2-\sqrt{\nu ^2+1}+1} \left(\nu ^2 z^4+2 \nu  z^3+2 \nu  z+1\right)}{-3 z^{-2} \ell  \left(\nu ^2 \chi ^2 z^4+2 \nu  \left(\chi ^2+1\right) z^3+\left(\chi ^2+3\right) z^2+1\right)^2}\,,\label{ji923} \\
 \phi &=&\left(\frac{\partial E}{\partial \mathcal{Q}}\right)_{\ell,\, S,\, V_2,\, c_2}=\sqrt{\frac{5}{2 \pi  \mathrm{G}_3}}\frac{g_3 \nu  \sqrt[4]{\nu ^2+1} \sqrt{\sqrt{\nu ^2+1}-1} \chi  z^3 (\nu  z+1)}{\ell  \left(\nu ^2 \chi ^2 z^4+2 \nu  \left(\chi ^2+1\right) z^3+\left(\chi ^2+3\right) z^2+1\right)}\,.  \label{jfi9p3e}
\end{eqnarray}
These respectively are conjugate to the entropy $\mathcal{S}$, thermodynamic volume $V_2$, central charge $c_2$, and electric charge $\mathcal{Q}$ of the  $\mathrm{DCFT}_2$. The full expression for $P_2$ is presented in Appendix \ref{joiew} and we here just expand it to the linear term of a small $\nu$ limit, i.e., in the small backreaction limit. Note that we have kept the backreaction parameter $\ell$ fixed; a fixed $\ell$ does not necessarily mean that the tension of the brane is fixed; cf. \eqref{fjioe3894}.  This seems to be the only choice, as we cannot obtain a  cohomogeneity thermodynamics if other quantities $\mathrm{G}_3,\,z$, or $\nu$ were fixed.  
 
 The first law on the $\mathrm{DCFT}_2$ reads
\begin{equation}
\mathrm{d}E=T\mathrm{d}\mathcal{S}-P_2 \mathrm{d}V_2+\mu_2 \mathrm{d}c_2+\phi \mathrm{d}\mathcal{Q}\,,
\end{equation}
where a work term $P_2 \mathrm{d}V_2$ indicates that the energy $E$ plays the role of the thermodynamic internal energy. Remarkably, different from the case of the brane perspective, the corresponding integral internal energy formula now contains the electric charge term, 
\begin{equation}\label{jfp2dm}
E=T \mathcal{S}+\mu_2 c_2+\frac{1}{2}\mathcal{Q}\phi\,.
\end{equation}
The energy formula is consistent with a scaling argument as $\mathrm{G}_4 E$ is a homogeneous function of $\mathrm{G}_3 S_{\mathrm{gen}},\,V_2/\mathrm{G}_3,\,,\mathrm{G}_3 c_2,\,$ and $\mathcal{Q}\,$. The absence of $V_2$  in the mass relation is due to that $V_2/\mathrm{G}_3$ is dimensionless. It is noticeable that $\mathrm{G}_3$ and $\nu$ are variable in the above configuration, which, according to \eqref{jfeorj39822}, means that the central charge $c_3$ in \eqref{jfeorj39822} on the brane is variable. If $\mathrm{G}_3$ were substituted with $\ell /\left(2 c_3 \sqrt{\nu ^2+1}\right)$ (cf. \eqref{jfeorj39822}), we can check that the results obtained above would be the same. In such a condition, $c_3$ does not enter the thermodynamic first law and the energy formula  as it is featured not for the  DCFT but for the holographic $\mathrm{CFT}_3$ on the brane as well as for the $\mathrm{BCFT}_3$.

\section{Closing remarks}

In this paper, we utilized the KR braneworld holography formulation to derive a quantum charged black hole on an $\mathrm{AdS}_3$ brane, starting from the $\mathrm{AdS}_4$ C-metric. We analyzed the backreaction effect from the $\mathrm{CFT}_3$ on the brane by calculating the holographic stress-energy tensor, revealing that the conformal symmetry property of the nonlinear electromagnetic field on the brane is lost. We derived the mass, generalized entropy, and electric charge, along with their respective conjugate thermodynamic quantities, for the quantum charged black hole. In the extended phase space, we scrutinized the thermodynamic first law and mass (energy) relations for classical and holographic thermodynamics of the quantum charged black hole from the perspectives of the pure bulk, brane, and boundary, adhering to the double holography prescription.

In the absence of charge, previous results indicated that the quantum black hole resembles the BTZ black hole under a specific parameter setup. However, the quantum charged black hole we derived is not analogous to the charged BTZ black hole \cite{Martinez:1999qi,Chan:1994qa} in any sense. As anticipated in \cite{Panella:2023lsi}, it resembles the RN-AdS black hole, primarily because we have a charge term proportional to $q^2/\bar{r}^2$. This is a consequence of the nonlinear nature of the gauge field on the brane. In our procedure to obtain this quantum charged black hole, we selected a brane at a specific position in the original $\mathrm{AdS}_4$ C-metric spacetime. This operation reduced the four-potential of the Maxwell field in the bulk to a three-potential. Consequently, the induced gravity on the brane is modified by higher curvature terms, and the induced electromagnetic field on the brane became nonlinear and coupled with the background spacetime curvature.

Note that throughout our study, we did not assign specific values to the parameter $\kappa$. The values of $\kappa=\pm 1,0$ correspond to different slicings of the brane. In the uncharged quBTZ scenario, both rotating and non-rotating BTZ black holes can be recovered for $\kappa=-1$, and there exist branches of black holes and black strings for specific values of the combination $\kappa x_1^2$. By conducting a similar parametric analysis, we believe that there could also be analogous branches of solutions for quantum charged objects within a limited mass range. These solutions include a branch with negative mass and a branch representing black strings for specific combinations of parameters $\{\kappa, x_1, q\}$; see \cite{Climent:2024nuj} for detailed analysis. Note that in our thermodynamic study of the black hole, $\kappa$ does not appear independently in the physical quantities. A more thorough analysis of the parameter space is necessary, particularly when our goal is to delve deeper into the thermodynamic phase transitions and the dynamical stability for the quantum charged black holes. Beyond the event horizon, the quantum charged black hole may  also have a Cauchy horizon. The stability of this Cauchy horizon is a topic that warrants further clarification. Additionally, studying the  quasinormal modes for the  quantum charged black hole on the brane could be a fruitful avenue for future research. This will provide us with a deeper understanding of the dynamical properties of the quantum charged black holes.

The study of the thermodynamics for the quantum charged black holes was inspired, in part, by the investigations in \cite{Frassino:2022zaz} on quBTZ black holes.  We have found that the holographic black hole chemistry \cite{Cong:2021fnf,Ahmed:2023snm} can be generalized to a doubly holographic scenario, leading to a number of interesting observations.  One is that the Smarr mass relation we derived in this paper, from the brane perspective, does not explicitly contain a charge term. This indicates that the quantum black hole owns a dimensionless gauge charge viewed from the brane perspective.  Another point is that, different from the result shown in \cite{Frassino:2022zaz}, the Euler energy relation \eqref{jfp2dm} we obtained from the pure boundary perspective does not contain the $\mu_3$ and $c_3$ terms. All quantities in this Euler energy relation are characterized by the DCFT. To achieve this, we fixed the backreaction parameter $\ell$, which, according to \eqref{dkjaoe834}, means that the acceleration of the bulk black hole is constant.

\acknowledgments

We thank Robie Hennigar for helpful discussions and the constructive revision suggestions from the referee. This work was supported by the Natural Sciences and Engineering Research Council of Canada and the National Natural Science Foundation of China (Grant Nos. 12365010, 12005080, and 12064018). M. Z. was also supported by the Chinese Scholarship Council Scholarship.

\appendix

\section{Wald entropy}\label{owi23i8}
For the Lagrangian \eqref{jeiuj3e389}, according to the formula given in \cite{Jacobson:1993vj},  the Wald entropy is
\begin{equation}
    S_W =\frac{1}{4\mathrm{G}_3}\int \mathrm{d}x \sqrt{q}\mathcal{W}\,,
\end{equation}
where 
\begin{equation}
    \mathcal{W}=1+\ell^2 (\left(a_1+2 a_2\right) R-a_1 q^{a b} R_{a b}+8 \pi \mathrm{G}_3\left(\left( - a_1-4 a_2\right) T-8 \pi \mathrm{G}_3 a_1 q^{a b} T_{a b}\right)+\mathcal{O}\left(\ell^4\right)\,.
\end{equation}
Here $a_1=-1,\,a_2=3/8$, $T$ is the trace of the stress-energy tensor $T_{ab}$, and $q_{ab}$ is the induced metric on a cross section of the black hole horizon. Considering \eqref{fejoi29} and \eqref{jf938p2}, we know that 
\begin{equation}
8 \pi \mathrm{G}_3\left\langle T_{a b}\right\rangle=R_{a b}-\frac{1}{2} g_{a b} R+\mathcal{O}\left(\ell^2\right)\,,
\end{equation}
in turn yielding \cite{Jacobson:1993vj}
\begin{equation}
S_W=\frac{1}{4 \mathrm{G}_3} \int \mathrm{d} x \sqrt{q}\left(1+\ell^2\left(2a_2 R+a_1 g_{\perp}^{a b} R_{a b}\right)\right)
\end{equation}
for  the Wald entropy of the quantum charged black hole up to $\mathcal{O}\left(\ell / \ell_3\right)^4$,  
where $g_{\perp}^{a b}=g^{ab}-q^{ab}$ is the metric normal to the black hole horizon, and $g^{ab}$ denotes \eqref{iroejreop1093}. Explicitly we obtain  
\begin{eqnarray}
    \frac{S_W}{S_{\mathrm{gen}}}&=&\frac{\nu ^3 \chi ^2 z^3 (\nu  z+1) (3 \nu  z+2)-\nu  \left(\nu +\nu ^2 z \left(2 z^2+3\right)-2 z\right)+2}{2 \sqrt{\nu ^2+1}}\nonumber\\&=&1-\nu ^2+\nu ^3 \left(\chi ^2 z^3-z^3-2 z\right)+\nu  z+\mathcal{O}\left(\nu^4\right)\,,\\
    \frac{S_W}{S_{\mathrm{cl}}}&=&\frac{\nu ^3 \chi ^2 z^3 (\nu  z+1) (3 \nu  z+2)-\nu  \left(\nu +\nu ^2 z \left(2 z^2+3\right)-2 z\right)+2}{2 \nu  z+2}\nonumber\\&=&1-\frac{\nu ^2}{2}+\nu ^3 \left(\chi ^2 z^3-z^3-z\right)+\mathcal{O}\left(\nu^4\right)\,.
\end{eqnarray}

\section{Explicit expressions of some thermodynamic quantities}\label{joiew}
The full expression of the chemical $\mu_3$ in \eqref{fj392p} is
\begin{equation}
\mu_3=-\frac{\left(\nu ^2+1\right) \left(X_1+X_2+X_3\right) z^2}{\ell_3 \left(\nu ^2+2 x+2\right) \left(z^2 \left((\chi +\nu  \chi  z)^2+2 \nu  z+3\right)+1\right)^2}\,,
\end{equation}
where
\begin{eqnarray}
 \nu_x&=&\sqrt{\nu ^2+1}\,,\nonumber\\
X_1&=&2 \nu  (\nu_x+2)+\nu ^2 \chi ^2 z^5 \left(\nu ^2+2 \left(\nu ^2-1\right) \nu_x-2\right)+z \left(\nu ^2 (2 \nu_x+5)-2 (\nu_x+1)\right)\,,\nonumber\\
X_2&=&2 \nu  z^4 \left(\nu ^2 \left(\chi ^2+2 \chi ^2 \nu_x+\nu_x+1\right)-2 \chi ^2 (\nu_x+1)\right)\,,\nonumber\\
X_3&=&z^3 \left(\nu ^2 \left(\chi ^2+2 \left(\chi ^2+3\right) \nu_x+7\right)-2 \left(\chi ^2-1\right) (\nu_x+1)\right)\,\nonumber.
\end{eqnarray}
The full expression of the pressure $P_2$ in \eqref{2j893qdp} is
\begin{equation}
P_2=-\frac{\left(Y_1+Y_2\right) z^2 \left(\nu ^2-\nu _x+1\right){}^{3/2} \nu _x}{2 \sqrt{2} \pi  \mathrm{G}_3 \nu ^2 Y_3^2 \ell  \left(\nu _x-1\right) \left(-2 \nu ^2+\nu _x-2\right)}\,,
\end{equation}
where
\begin{eqnarray}
Y_1&=&-\nu ^2+4 \nu _x+4 \nu ^5 \chi ^2 z^5 \nu _x-2 \nu  z \left(\nu ^2-4 \nu _x+4\right)-4\,,\nonumber\\
Y_2&=&\nu ^4 z^4 \left(8 \chi ^2 \nu _x-4 \nu _x+3\right)+2 \nu ^3 z^3 \left(2 \chi ^2 \nu _x-4 \nu _x+3\right)\,,\nonumber\\
Y_3&=&\nu ^2 \chi ^2 z^4+2 \nu  \left(\chi ^2+1\right) z^3+\left(\chi ^2+3\right) z^2+1\,.\nonumber
\end{eqnarray}

\bibliographystyle{JHEP}
\bibliography{biblio.bib}

\end{document}